\documentclass[reprint,superscriptaddress,nofootinbib,twocolumn,amsmath,amssymb,aps,prl]{revtex4-2}

\usepackage{graphicx}
\usepackage{svg}
\usepackage{dcolumn}
\usepackage{bm}
\usepackage{xcolor}
\usepackage{braket}
\usepackage[colorlinks]{hyperref}
\usepackage{bm}
\usepackage{bbold}

\newcommand{\pare}[1]{\left( #1 \right)}
\newcommand{\ave}[1]{\langle #1 \rangle}
\newcommand{\sinc}{\text{sinc}}
\newcommand{\cor}[1]{\left[ #1 \right]}
\newcommand{\revA}[1]{\textcolor{black}{#1}}
\newcommand{\revB}[1]{\textcolor{black}{#1}}
\newcommand{\revC}[1]{\textcolor{black}{#1}}
\newcommand{\editor}[1]{\textcolor{black}{#1}}

\begin{document}

\preprint{APS/123-QED}

\title{Multiphoton Quantum van Cittert–Zernike Theorem}

\author{Ashe Miller}
\affiliation{Quantum Photonics Laboratory, Department of Physics \& Astronomy, Louisiana State University, Baton Rouge, LA 70803, USA}
\author{Chenglong You}
\affiliation{Quantum Photonics Laboratory, Department of Physics \& Astronomy, Louisiana State University, Baton Rouge, LA 70803, USA}

\author{Roberto de J. Le\'on-Montiel}
\email{roberto.leon@nucleares.unam.mx}
\affiliation{Instituto de Ciencias Nucleares, Universidad Nacional Aut\'onoma de M\'exico, Apartado Postal 70-543, 04510 Cd. Mx., M\'exico}

\author{Omar S. Maga\~na-Loaiza}
\affiliation{Quantum Photonics Laboratory, Department of Physics \& Astronomy, Louisiana State University, Baton Rouge, LA 70803, USA}

\date{\today}

\begin{abstract}
Recent progress on quantum state engineering has enabled the preparation of quantum photonic systems comprising multiple interacting particles. Interestingly, multiphoton quantum systems can host many complex forms of interference and scattering processes that are essential to perform operations that are intractable on classical systems. Unfortunately, the quantum coherence properties of multiphoton systems degrade upon propagation leading to undesired quantum-to-classical transitions. Furthermore, the manipulation of multiphoton quantum systems requires of nonlinear interactions at the few-photon level.  Here, we introduce the quantum van Cittert-Zernike theorem to describe the scattering and interference effects of propagating multiphoton systems. This fundamental theorem demonstrates that the quantum statistical fluctuations, which define the nature of diverse light sources, can be modified upon propagation in the absence of light-matter interactions. The generality of our formalism unveils the conditions under which the evolution of multiphoton systems can lead to surprising classical-to-quantum transitions. Specifically, we show that the implementation of conditional measurements may enable the all-optical preparation of multiphoton systems with attenuated quantum statistics below the shot-noise limit. Remarkably, this effect had not been discussed before and cannot be explained through the classical theory of optical coherence. As such, our work opens new paradigms within the established field of quantum coherence.

\end{abstract}

\maketitle


The van Cittert-Zernike theorem constitutes one of the pillars of optical physics \cite{Cittert:34,ZERNIKE:38}. \revB{As such, this fundamental theorem provides the  formalism to describe the modification of the coherence properties of optical fields upon propagation \cite{Cittert:34,ZERNIKE:38,born2013principles,Wolf:54}.} Over the last decades, extensive investigations have been conducted to explore the evolution of spatial, temporal, spectral, and polarization coherence of diverse families of optical beams \cite{Dorrer:04, gori2000use, cai2020, Cai2012}. In the context of classical optics, the investigation of the van Cittert-Zernike theorem led to the development of schemes for optical sensing, metrology, and astronomical interferometry \cite{Carozzi:09, Batarseh:18, Barakat:2000}. Nowadays, there has been interest in exploring the implications of the van Cittert-Zernike theorem for quantum mechanical systems \revB{\cite{PhysRevA.54.4473,Saleh05PRL,Fabre_2017, Howard2019,Barrachina2020,PhysRevLett.123.070504}}. Recent efforts have been devoted to study the evolution of the properties of spatial coherence of biphoton systems \revB{\cite{PhysRevA.54.4473,Saleh05PRL, Howard2019, PhysRevA.95.063836, PhysRevA.99.053831,PhysRevLett.123.070504}. Specifically, the van Cittert-Zernike theorem has been extended to analyze the spatial entanglement between a pair of photons generated by parametric down conversion \cite{PhysRevA.54.4473,Saleh05PRL,Qian:18,Eberly_2016}.} The description of the evolution of spatial coherence and entanglement of propagating photons turned out essential for quantum metrology, spectroscopy, imaging, and lithography \revB{\cite{PhysRevA.54.4473,Howard2019,PhysRevLett.123.070504, bhusal2021smart, de_J_Le_n_Montiel_2013, Saleh05PRL, you2021scalable, obrien_photonic_2009,Wen:13}}. Nevertheless, previous research has not explored the evolution of the excitation mode of the field that establishes the quantum statistical properties and the nature of light sources \cite{PhysRevA.54.4473,Saleh05PRL,Fabre_2017, Howard2019,Barrachina2020,PhysRevLett.123.070504,PhysRevA.54.4473,Saleh05PRL, Howard2019, PhysRevA.95.063836, PhysRevA.99.053831,PhysRevLett.123.070504,PhysRevA.54.4473,Saleh05PRL,Qian:18,Eberly_2016, PhysRevA.54.4473,Howard2019,PhysRevLett.123.070504, bhusal2021smart, de_J_Le_n_Montiel_2013, Saleh05PRL, you2021scalable, obrien_photonic_2009,Wen:13}.

There has been important progress on the preparation of multiphoton systems with quantum mechanical properties \cite{magana2019multiphoton, DELLANNO200653,OLSEN2002373,munoz_emitters_2014}. The interest in these systems resides in the complex interference and scattering effects that they can host \revB{\cite{aspuru-guzik_photonic_2012, obrien_photonic_2009, You2020plasmonics,munoz_emitters_2014}}. Remarkably, these fundamental processes define the statistical fluctuations of photons that establish the nature of light sources \cite{Mandel:79, you2020identification,magana2019multiphoton, DELLANNO200653,mandel1995optical,OLSEN2002373}. Furthermore, these quantum fluctuations are associated to distinct excitation modes of the electromagnetic field that determine the quantum coherence of a light field \cite{Mandel:79,mandel1995optical}. In the context of quantum information processing, the interference and scattering among photons have enormous potential to perform operations that are intractable on classical systems \cite{aspuru-guzik_photonic_2012,obrien_photonic_2009}. However, the manipulation of multiphoton systems requires complex light-matter interactions that are hard to achieve at the few-photon level \cite{venkataraman_phase_2013,DELLANNO200653}. Indeed, it has been assumed that light-matter interactions are needed to modify the excitation mode of an optical field \cite{you2021observation,tame_mix_2021}. These challenges have motivated interest in linear optical circuits for random walks, boson sampling, and quantum computing \cite{knill2001, RevModPhys.79.135,aspuru-guzik_photonic_2012}. Moreover, the interaction of multiphoton quantum systems with the environment leads to the degradation of their nonclassical properties upon propagation \cite{Yu_Sudden_2009, You2020plasmonics}. Indeed, quantum-to-classical transitions are unavoidable in nonclassical systems interacting with realistic environments \cite{Yu_Sudden_2009}.  These vulnerabilities have prevented the use of nonclassical states of light for the sensing of small physical parameters with sensitivities that surpass the shot-noise limit \cite{Polino_Photonic_2020, you2021scalable}. The possibility of using nonclassical multiphoton states to demonstrate scalable quantum sensing has constituted one of the main goals of quantum optics for many decades \cite{Polino_Photonic_2020}.

In contrast to well-established paradigms in the field of quantum optics \cite{PhysRevA.54.4473,Saleh05PRL,Fabre_2017, Howard2019,Barrachina2020,PhysRevLett.123.070504,PhysRevA.54.4473,Saleh05PRL, Howard2019, PhysRevA.95.063836, PhysRevA.99.053831,PhysRevLett.123.070504,PhysRevA.54.4473,Saleh05PRL,Qian:18,Eberly_2016, PhysRevA.54.4473,Howard2019,PhysRevLett.123.070504, bhusal2021smart, de_J_Le_n_Montiel_2013, Saleh05PRL, you2021scalable, obrien_photonic_2009,Wen:13}, our work demonstrates that the quantum statistical fluctuations of multiphoton light fields can be modified upon propagation in the absence of light-matter interactions. We introduce the quantum van Cittert-Zernike theorem to describe the underlying scattering effects that give rise to classical-to-quantum transitions in multiphoton systems. Remarkably, our work unveils the conditions under which sub-shot-noise quantum fluctuations can be extracted from thermal light sources. These effects remained elusive for multiple decades due to the limitations of the classical theory of optical coherence to describe multiphoton scattering \cite{PhysRevA.54.4473,Saleh05PRL,Fabre_2017, Howard2019,Barrachina2020,PhysRevLett.123.070504,PhysRevA.54.4473,Saleh05PRL, Howard2019, PhysRevA.95.063836, PhysRevA.99.053831,PhysRevLett.123.070504,PhysRevA.54.4473,Saleh05PRL,Qian:18,Eberly_2016, PhysRevA.54.4473,Howard2019,PhysRevLett.123.070504, bhusal2021smart, de_J_Le_n_Montiel_2013, Saleh05PRL, you2021scalable, obrien_photonic_2009,Wen:13}. This stimulated the idea that the quantum statistical fluctuations of light fields were not affected upon propagation. Our work provides an all-optical alternative to prepare multiphoton systems with sub-Poissonian statistics. Previously, similar functionalities have been demonstrated in nonlinear optical systems, photonics lattices, plasmonic systems, and Bose-Einstein condensates \cite{magana2019multiphoton, you2021observation, kondakci_photonic_2015,OLSEN2002373}. 

\begin{figure}[t!]
 \includegraphics[width=0.9\linewidth]{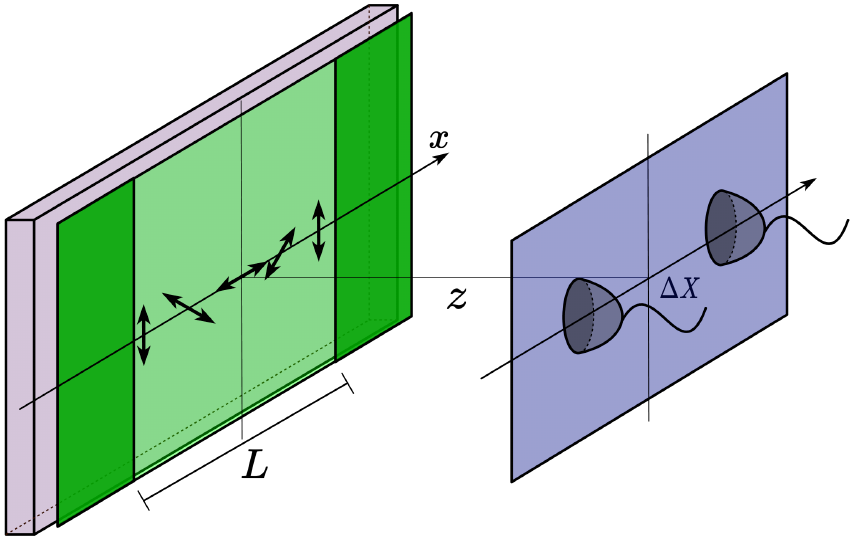}
 \caption{The proposed setup for investigating the multiphoton quantum van Cittert-Zernike theorem. We consider an incoherent, unpolarized beam interacting with a polarization grating of length $L$ at $z=0$. After interacting with the grating, the beam propagates a distance of $z$ onto the measurement plane, where two point detectors  are placed $\Delta X$ apart.} 
 \label{fig:Setup}
\end{figure}

We demonstrate the multiphoton quantum van-Cittert Zernike theorem by extending the work of Gori et al. \cite{gori2000use} to two-mode correlations using the setup in Fig. \ref{fig:Setup}. In general, each mode can host a multiphoton system with an arbitrary number of photons. We consider a thermal, spatially incoherent, unpolarized beam that interacts with a polarization grating. This grating modifies the polarization of the thermal beam at different transverse spatial locations $x$ according to ${\pi x}/{L}$. Here, $L$ represents the length of the grating. The thermal beam propagates to the far-field, where it is measured by two point detectors \cite{magana-loaiza-2016,Shih2009PRA,Chekhova08PRA}. We then post-select on the intensity measurements made by these detectors to quantify the correlations between different modes of the beam. 

The multiphoton quantum van Cittert-Zernike theorem can be demonstrated for any incoherent, unpolarized state, the simplest of which is an unpolarized two-mode state \cite{soderholm2001unpolarized}. The two-mode state can be produced by a source emitting a series of spatially independent photons with either horizontal (H) or vertical (V) polarization, giving an initial state \cite{magana2019multiphoton,wei2005synthesizing}
\begin{equation}
\begin{aligned}
\hat{\rho}&=\hat{\rho}_{1} \otimes \hat{\rho}_{2}\\
&= \frac{1}{4}\left(\left|H\right\rangle_{1}\left|H\right\rangle_{2}\left\langle H\right|_{1}\left\langle H\right|_{2}\right.+\left|H\right\rangle_{1}\left|V\right\rangle_{2}\left\langle H\right|_{1}\left\langle V\right|_{2} \\
&\quad+\left|V\right\rangle_{1}\left|H\right\rangle_{2}\left\langle V\right|_{1}\left\langle H\right|_{2}\left.+\left|V\right\rangle_{1}\left|V\right\rangle_{2}\left\langle V\right|_{1}\left\langle V\right|_{2}\right),
\end{aligned}
\label{eqn:state}
\end{equation}
where the subscripts denote the mode. For simplicity, we begin by considering the case where a single photon is emitted in each mode.

We can find the state immediately after the polarization grating shown in Fig. \ref{fig:Setup} to be
\begin{equation}
\begin{aligned}
  \hat{\rho}_{\text{pol}}=\hat{P}\pare{x_1}\hat{\rho}_1\hat{P}\pare{x_2}\otimes\hat{P}\pare{x_3}\hat{\rho}_2\hat{P}\pare{x_4},
\end{aligned}
\label{eqn:pol}
\end{equation}
where $\hat{P}(x)$ is the projective measurement given by
\begin{equation}
\begin{aligned}
  \hat{P}\pare{x}=\begin{bmatrix}
  \cos^2\pare{\frac{\pi x}{L}} && \cos\pare{\frac{\pi x}{L}}\sin\pare{\frac{\pi x}{L}} \\
  \cos\pare{\frac{\pi x}{L}}\sin\pare{\frac{\pi x}{L}} &&\sin^2\pare{\frac{\pi x}{L}}
  \end{bmatrix}.
\end{aligned}
\label{eqn:Prop} 
\end{equation}

For ease of calculation, we utilize the Heisenberg picture, back-propagating the detector operators to the polarization grating. The point detector is modeled by  $\hat{O}_{j,k,z}(X)=\hat{a}^{\dag}_{j,z}(X)\hat{a}_{k,z}(X)$, where $z$ is the distance between the grating and the measurement plane. The ladder operator $\hat{a}_z(X)$ is defined as
\begin{equation}
\begin{aligned}
  \hat{a}_{j,z}(X)= \int^{\frac{L}{2}}_{-\frac{L}{2}} dx \hat{a}_{j,0}\pare{x}\text{Exp}[-\frac{2\pi i}{z\lambda}xX],
\end{aligned}
\label{eqn:ladOp}
\end{equation}
where $X$ is the position of the detector on the measurement plane, $\lambda$ is the wavelength of the beam and $j,k$ is the polarization of the operator. Eq. (\ref{eqn:ladOp}) describes  the contribution of each point on the polarization grating plane to the detection measurement. Since we wish to keep the information of each interaction on the screen, we choose to calculate the four-point auto covariance by \cite{gureyev2017van, perez2017two}
\begin{equation}\label{eqn:Coh1}
\begin{aligned}
  G_{jklm}^{\pare{2}}&\pare{\mathbf{X},z}=\\&\text{Tr}[\hat{\rho}_{\text{pol}}\hat{a}^{\dag}_{j,z}\pare{X_1}\hat{a}_{k,z}\pare{X_2}\hat{a}^{\dag}_{l,z}\pare{X_3}\hat{a}_{m,z}\pare{X_4}],
\end{aligned}
\end{equation}
where $\mathbf{X}=\left[X_1,X_2,X_3,X_4\right]$, allowing for the measurement of a post-selected coherence. We then set $X_2=X_1$ and $X_4=X_3$, since we are working with two point detectors. We allow the operators of the two detectors to commute, recovering the well-known expression for second-order coherence \cite{gerry2005introductory}. The second-order coherence of any post-selected measurement is then found to be
\begin{equation}\label{eqn:2ndOrd}
\begin{aligned}
&G_{j k l m}^{(2)}(\mathbf{X},z)=\int d x_{1} \int d x_{2} \int d x_{3} \int d x_{4}\\
&\times C_{j k l m}(\boldsymbol{x}) F(\boldsymbol{x}, \mathbf{X},z) [\delta\left(x_{1}-x_{2}\right) \delta\left(x_{3}-x_{4}\right)\\
&+\delta\left(x_{1}-x_{4}\right) \delta\left(x_{3}-x_{2}\right)],
\end{aligned}
\end{equation}
where the limits of integration for each integral is $-L/2$ to $L/2$, $\boldsymbol{x}=[x_1,x_2,x_3,x_4]$, $C_{jklm}\pare{\bm{x}}$ is the coefficient of the $\ket{j}_1\ket{k}_2\bra{l}_1\bra{m}_2$ element of the density matrix $\hat{\rho}_{\text{pol}}$ in Eq. (\ref{eqn:pol}), and $j,k,l,m \in \{H,V\}$. Furthermore, $F\pare{\bm{x},\mathbf{X},z}$ is given as
\begin{equation}
\begin{aligned}
  F\pare{\bm{x},\mathbf{X},z}=\text{Exp}[\frac{2\pi i}{\lambda z}\pare{X_4x_4-X_3x_3+X_2x_2-X_1x_1}].
\end{aligned}
\label{eqn:prop}
\end{equation}

\begin{figure}
 \includegraphics[width=0.95\linewidth]{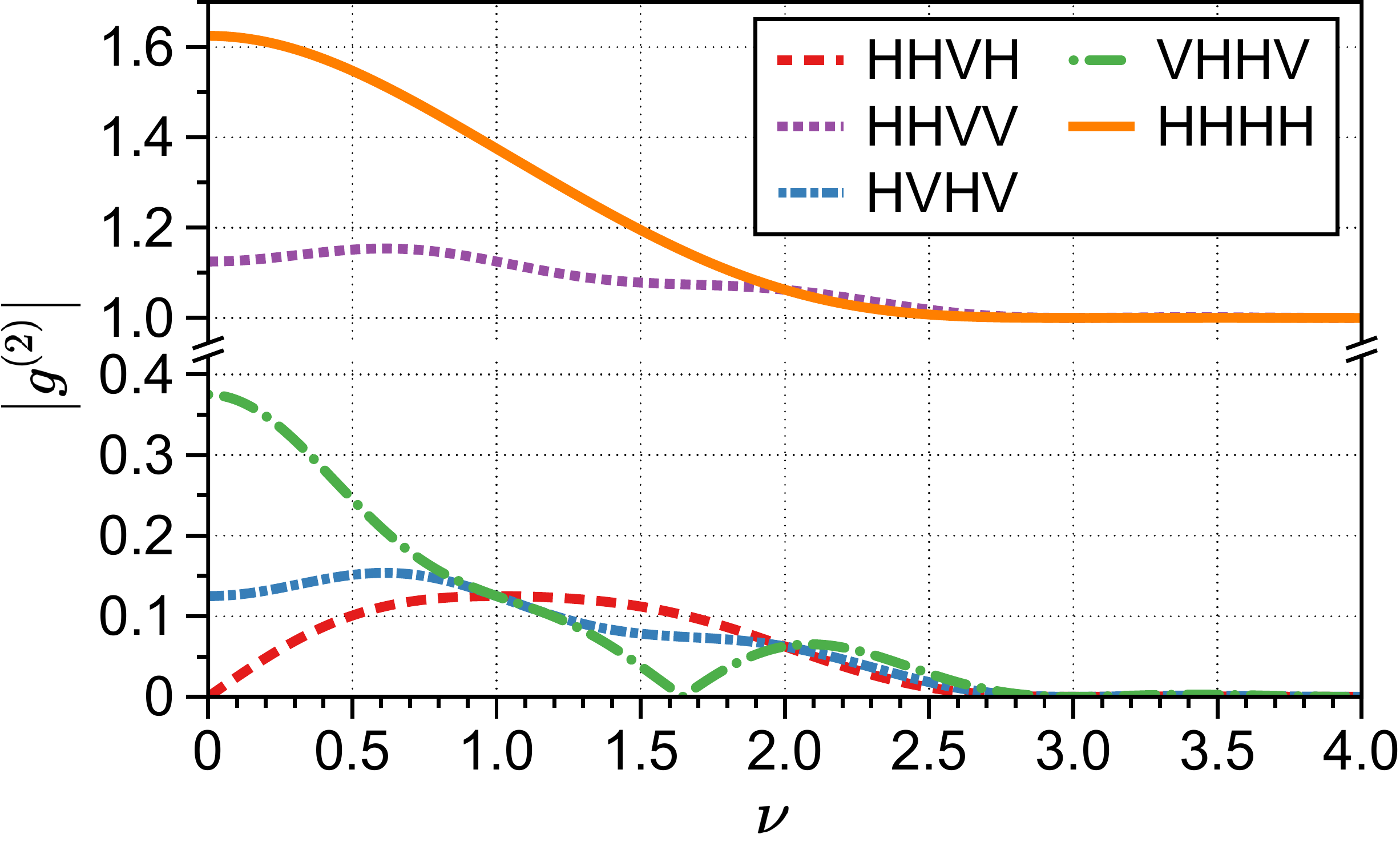}
 \caption{The second-order coherence for various post-selected measurements \revA{in the far field}. The $x$-axis is how the $g^{(2)}$ changes as a function of $\nu=L\Delta X/\pare{\lambda z}$ while keeping $L$, $\lambda$ and $z$ fixed. As the detectors move further apart, the spatial correlations created by the polarization grating decrease until they diminish entirely at $\nu\approx2.7$. In addition, certain post-selected measurements allow us to quantify the coherence between two fields that possess sub-Poissonian statistics, suggesting the possibility of sub-shot noise measurements. Note, these measurements can be also performed using quantum state tomography \cite{cramer2010efficient}.} 
 \label{fig:Components}
\end{figure}

We set $X_2=X_1$ and $X_4=X_3$, which properly describes the two point detectors allowing Eq. (\ref{eqn:2ndOrd}) to become a 2D Fourier transform \cite{gori2000use}. By observing  Eq. (\ref{eqn:2ndOrd}), it is important to note that there are two spatial correlations that contribute to the coherence at the measurement plane. One is the correlation of a photon with itself which existed prior to interacting with the polarizer, while the other is the spatial correlation gained by two photon scattering. It is important to note that the key difference between our formalism and the classical formalism is the existence of the scattering term as discussed in the supplementary material. Due to the nature of projective measurements in Eq. (\ref{eqn:pol}), the density matrix $\hat{\rho}_{\text{pol}}$ will no longer be diagonal in the horizontal-vertical basis, allowing for the beam to temporarily gain and lose polarization coherence \cite{gori2000use}. The self-coherence of a photon results in the minimum coherence throughout all measurements in the far-field. The correlations from two photon scattering sets the maximum coherence and determines how it changes with the distance between the detectors.

To extend the description of a two-mode system comprising of two photons \revB{to a multiphoton picture capable of handling any state, we need to propagate a value other than the photon statistics.} Attempting to propagate a multiphoton field under the Schrodinger and Heisenberg pictures becomes computationally hard, scaling on the order of $O(2^n n!)$ where $n$ represents the number of photons \cite{10.1145/1993636.1993682}. As a result, we propagate multiphoton states using a quantum version of the beam coherence polarization (BCP) matrix \cite{gori1998beam,gori2000use}.  This formalism allows us to estimate the evolution of the four-point-correlation matrix, reducing the total elements of interest. Consequently, the two-photon calculation represents the simplest case that the BCP matrix can handle and is in agreement with the general multiphoton picture.  We begin by defining the BCP matrix as
\textcolor{black}{\begin{multline}
\ave{\hat{J}\pare{X_1,X_2,z}}=\\
\begin{bmatrix}
\ave{\hat{E}_H^\dag\pare{X_1,z}\hat{E}_H\pare{X_2,z}} && \ave{\hat{E}_H^\dag\pare{X_1,z}\hat{E}_V\pare{X_2,z}}\\
\ave{\hat{E}_V^\dag\pare{X_1,z}\hat{E}_H\pare{X_2,z}} && \ave{\hat{E}_V^\dag\pare{X_1,z}\hat{E}_V\pare{X_2,z}}
\end{bmatrix}.
\end{multline}}
Here, the angle brackets denote the ensemble average, whereas the quantities $\hat{E}_{\alpha}^\dag\pare{X,z}$ and $\hat{E}_{\alpha}\pare{X,z}$ represent the negative- and positive-frequency components of the $\alpha$-polarized (with $\alpha = H,V$) field-operator at the space-time point $(X,z;t)$, respectively. We can then propagate the BCP matrix through the grating and to the measurement plane, by considering an initial BCP matrix of the form: $I_2\delta\pare{X_1-X_2}$ \cite{gori1998beam}. The details of the calculation can be found in the supplementary material. Upon reaching the measurement plane we can find the second-order coherence matrix given by \cite{Pires:21}
\begin{equation} \label{eqn:Coh2}
\begin{aligned}
 \mathbf{G}^{(2)}(\bm{X},z)=\ave{\hat{J}\pare{X_1,X_2,z}\otimes\hat{J}\pare{X_3,X_4,z}}.
\end{aligned}
\end{equation}
Each element of the $\mathbf{G}^{(2)}$ matrix is a post-selected coherence matching each combination of polarizations shown in Eqs. (\ref{eqn:Coh1})-(\ref{eqn:2ndOrd}). As shown in the supplementary material, the result obtained is equivalent to the approach described in Eqs. (\ref{eqn:state})-(\ref{eqn:prop}).


In order to demonstrate the results of our calculation, we first look at the second-order coherence of the horizontal mode in the far-field. By normalizing either Eq. (\ref{eqn:Coh1}) or the matrix element of Eq. (\ref{eqn:Coh2}), we find the coherence of the horizontal mode to be
\begin{widetext}
\begin{equation}
\begin{split}
  g^{(2)}_{\text{HHHH}}&(\nu)=1+\frac{1}{16}\sinc^2\pare{2-\nu}+\frac{5}{8}\sinc^2\pare{\nu}+\frac{1}{16}\sinc^2\pare{2+\nu}
  +\frac{1}{4}\sinc\pare{2-\nu}\sinc\pare{1-\nu}+\frac{3}{8}\sinc^2\pare{1-\nu}\\
  &+\frac{1}{4}\sinc\pare{1+\nu}\pare{\sinc\pare{2+\nu}+\sinc\pare{1-\nu}}+\frac{3}{8}\sinc^2\pare{1+\nu}\\
  &+\frac{1}{8}\sinc\pare{\nu}\pare{\sinc\pare{2-\nu}+\sinc\pare{2+\nu}+6\sinc\pare{1-\nu}+6\sinc\pare{1+\nu}},
\end{split}
\label{eqn:g2HHHH}
\end{equation}
\end{widetext}
where $g^{(2)}_{jklm}$ is the normalized second-order coherence. Here $\sinc(\nu)=\sin(\pi \nu)/\pare{\pi \nu}$ and $\nu=L\Delta X/\pare{\lambda z}$. Therefore, $g^{(2)}_{jklm}$ depends on the distance between the detectors $\Delta X = X_1 - X_2$, the length of the polarization grating $L$, the wavelength $\lambda$, and the distance in the far field $z$. The same holds true for all other ${g}^{(2)}_{jklm}$, where each expression can be found in the supplementary material. \revA{Since Eq. (\ref{eqn:g2HHHH}) applies to all incoherent unpolarized states, we will perform the analysis for two mode thermal states, as the statistical properties are well studied \cite{mandel1995optical,gerry2005introductory,you2021observation,you2020identification,you2021observation}}.

\begin{figure*}
 \includegraphics[width=0.95\linewidth]{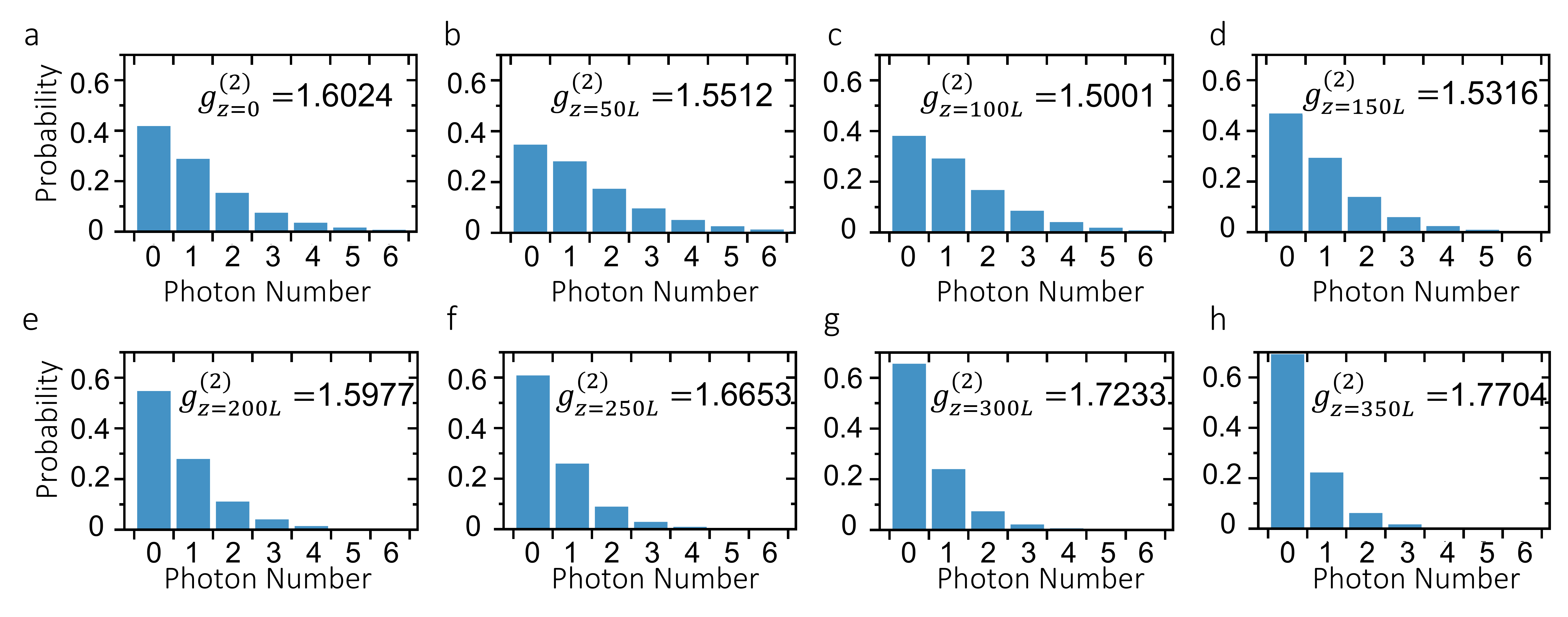}
 \caption{ The modification of the photon-number distribution and quantum coherence of a thermal multiphoton system upon propagation. In this case, the multiphoton system comprises a mixture of single-mode photons with either vertical or horizontal polarization. We assumed a single photon-number-resolving detector placed at different propagation distances: \textbf{a} $z=0$, \textbf{b} $z=50L$, \textbf{c} $z=100L$, \textbf{d} $z=150L$, \textbf{e} $z=200L$, \textbf{f} $z=250L$, \textbf{g} $z=300L$, \textbf{h} $z=350L$. In the transverse plane, the photon-number-resolving detector is placed at $X=0.4 L$. } 
 \label{fig:Components2}
\end{figure*}

As shown in Fig. \ref{fig:Components}, increasing the separation $\Delta X$ of the detectors causes the correlations to gradually decrease. Once $\nu\approx 2.7$, the detectors become uncorrelated. We note that $g^{(2)}(\nu)=1, \nu\neq 0$ represents an uncorrelated measurement since this can only be true when the two spatial modes become separable. In addition, when one of the two measured modes is no longer contributing to the measurement we get a $g^{(2)}(\nu)=0$. Interestingly, by fixing the distance $\Delta X$ between the two detectors, we can increase the correlations by moving the measurement plane further into the far-field. This is equivalent to decreasing $\nu$, causing correlations to increase to a possible maximum value of $g^{(2)}(0)=1.62$, suggesting an increase in bunching \cite{mandel1995optical}. By measuring $g^{(2)}_{\text{grating}}(0)$ immediately after the polarization grating at $x=0$, a horizontally polarized beam is measured with a $g^{(2)}_{\text{grating}}(0)=2$. We note the theory we presented only applies to the far-field, therefore these two values do not contradict each other. While the exact transition between the near and the far-fields are beyond the scope of the paper, we note that the horizontal mode along the central axis becomes more coherent as it propagates to the far-field, as predicted by the van Cittert-Zernike theorem \cite{gureyev2017van, Fabre_2017}. 

Setting one detector to measure the vertical mode and the other detector the horizontal mode, given by $g^{(2)}_{\text{HHVV}}$ in Fig. \ref{fig:Components}, we can measure the coherence between the horizontal and vertical mode. This post-selective measurement results in a different effect from when we only measured only the horizontal mode. Placing the detectors immediately after the polarization grating at $x=0$, we measure $g^{(2)}_{\text{grating}}(0)=0$ since there is no vertically polarized mode. However, we measure $g^{(2)}(0)\approx 1.1$ when the beam is propagated into the far-field. In this case, the polarization grating leads to the thermalization of the beam \cite{you2021observation}. This can be verified by removing the polarization grating and repeating the measurement giving $g^{(2)}_{\text{initial}}(0)=1$ since the two modes are completely uncorrelated.

The measurements of $g^{(2)}_{\text{HHHH}}$ and $g^{(2)}_{\text{HHVV}}$ can be performed using point detectors, however we predict more interesting effects that can be observed through the full characterization of the field. This information can be obtained through quantum state tomography \cite{cramer2010efficient}. We find that the second-order coherence $g^{(2)}_{\text{VHHV}}$, $g^{(2)}_{\text{HHVH}}$ and $g^{(2)}_{\text{HVHV}}$ is below one suggesting sub-Poissonian statistics, which potentially allows for sub-shot-noise measurement \cite{agarwal2012quantum}. It is important to note that while $g^{(2)}_{\text{HHVH}}(0)<g^{(2)}_{\text{HHVH}}(\nu)$  would be indicative of anti-bunching \cite{agarwal2012quantum}, however the term is imaginary. This feature is found using the BCP matrix approach, therefore it is true for all unpolarized incoherent fields. The sub-Poissonian statistics were achieved only with the use of post-selection without nonlinear interactions  \cite{magana2019multiphoton,you2021scalable,obrien_photonic_2009}. \revA{Our theoretical formalism
unveils the possibility of modifying the quantum
statistics of the electromagnetic field in free space without light-matter interactions \cite{you2021observation,DELLANNO200653,kondakci_photonic_2015,FollingNature}. Remarkably, the possibility of reducing the quantum fluctuations of photons below the shot-noise limit has represented one of the main goals of quantum engineers for many decades \cite{you2021scalable}.} Another interesting feature is that the $g^{(2)}_{\text{VHHV}}$ decays and resurrects  at $\nu\approx 1.6$ before decaying again. \revA{The fact that these elements are nonzero contrasts classical analyses where off-diagonal correlations of the system are  zero since the system is unpolarized in the far-field \cite{gori2000use,soderholm2001unpolarized}. In fact the classical analysis fails to describe any emergent phenomena shown since the final density matrix is the identity matrix (see supplementary materials). This new quantum degree of polarization is likely caused by the two photon scattering induced by the polarization grating. Furthermore, the creation of nonzero off diagonal elements suggests that we induced quantum correlations in our system. Ultimately, features noted in the above paragraph represent a mechanism to induce  classical to quantum dynamics.}

The sub-Possonian statistics are exclusive to unpolarized systems.  Returning to Eq. (\ref{eqn:2ndOrd}) we have that there are two correlations contributing to the final coherence, one from the photons self-coherence that existed prior to interaction with the screen and another coherence term that comes from photon scattering.  For unpolarized states there is no initial correlations in the off-diagonal elements of the density matrix, since by definition the off diagonal elements are zero \cite{soderholm2001unpolarized}.  This results in the first term of Eq. (\ref{eqn:2ndOrd}) to be zero for all off diagonal elements.  As noted above, this term sets the minimum value of the coherence measurement to zero, allowing for the measurement of sub-Poissonian statistics. 

Finally, we would like to highlight the fact that the quantum statistical properties of multiphoton systems can change upon propagation \revC{without light-matter interactions}
due to the scattering of their constituent single-mode photons carrying different polarizations.  This effect is quantified through the second-order quantum coherence $g^{(2)}(\tau=0)$ defined as $g^{(2)}(\tau=0)=1+\left(\left\langle(\Delta \hat{n})^{2}\right\rangle-\langle\hat{n}\rangle\right) /\langle\hat{n}\rangle^{2}$ \cite{Mandel:79, mandel1995optical}. In this case, the averaged quantities in $g^{(2)}(\tau=0)$ are obtained through the density matrix of the system's state, as described in Eq. (\ref{eqn:state}),
at different spatial coordinates ($\boldsymbol{X},z$). In Figure \ref{fig:Components2}, we report the photon-number distribution of the combined vertical-horizontal multiphoton field. In this case, a single photon-number-resolving detector was placed at $X=0.4 L$ \cite{you2020identification}. Note that by selecting the proper propagation distance $z$, one could, in principle, generate on-demand multiphoton systems with sub-Poissonian or Poissonian statistics \cite{magana2019multiphoton,you2021observation}, see the supplementary materials for details on the combined-field photon-number distribution calculation. As indicated in Figs. \ref{fig:Components} and  \ref{fig:Components2}, the evolution of quantum coherence  upon propagation lies at the heart of the quantum van Cittert-Zernike theorem for multiphoton systems.

In conclusion, we have investigated new mechanisms to control nonclassical coherence of multiphoton systems. We describe these interactions using a quantum version of the van Cittert-Zernike theorem. Specifically, by considering a polarization grating together with conditional measurements, we show that it is possible to control the quantum coherence of multiphoton systems. Moreover, we unveiled the possibility of producing multiphoton systems with sub-Poissonian statistics without light-matter interactions \cite{you2021observation,DELLANNO200653,kondakci_photonic_2015,FollingNature}. This possibility had not been discussed before,
and cannot be explained through the classical theory of optical coherence \cite{Wolf:54}. Thus, our work demonstrates that the multiphoton quantum van Cittert-Zernike theorem will have important implications for describing the evolution of the properties of quantum coherence of many-body bosonic systems \cite{You2020plasmonics,DELLANNO200653}. As such, our findings could offer alternatives to creating novel states of light by controlling the collective interactions of many single-photon emitters \cite{munoz_emitters_2014}.

\bibliography{Bibliography}

\begin{thebibliography}{59}%
\makeatletter
\providecommand \@ifxundefined [1]{%
 \@ifx{#1\undefined}
}%
\providecommand \@ifnum [1]{%
 \ifnum #1\expandafter \@firstoftwo
 \else \expandafter \@secondoftwo
 \fi
}%
\providecommand \@ifx [1]{%
 \ifx #1\expandafter \@firstoftwo
 \else \expandafter \@secondoftwo
 \fi
}%
\providecommand \natexlab [1]{#1}%
\providecommand \enquote  [1]{``#1''}%
\providecommand \bibnamefont  [1]{#1}%
\providecommand \bibfnamefont [1]{#1}%
\providecommand \citenamefont [1]{#1}%
\providecommand \href@noop [0]{\@secondoftwo}%
\providecommand \href [0]{\begingroup \@sanitize@url \@href}%
\providecommand \@href[1]{\@@startlink{#1}\@@href}%
\providecommand \@@href[1]{\endgroup#1\@@endlink}%
\providecommand \@sanitize@url [0]{\catcode `\\12\catcode `\$12\catcode
  `\&12\catcode `\#12\catcode `\^12\catcode `\_12\catcode `\%12\relax}%
\providecommand \@@startlink[1]{}%
\providecommand \@@endlink[0]{}%
\providecommand \url  [0]{\begingroup\@sanitize@url \@url }%
\providecommand \@url [1]{\endgroup\@href {#1}{\urlprefix }}%
\providecommand \urlprefix  [0]{URL }%
\providecommand \Eprint [0]{\href }%
\providecommand \doibase [0]{https://doi.org/}%
\providecommand \selectlanguage [0]{\@gobble}%
\providecommand \bibinfo  [0]{\@secondoftwo}%
\providecommand \bibfield  [0]{\@secondoftwo}%
\providecommand \translation [1]{[#1]}%
\providecommand \BibitemOpen [0]{}%
\providecommand \bibitemStop [0]{}%
\providecommand \bibitemNoStop [0]{.\EOS\space}%
\providecommand \EOS [0]{\spacefactor3000\relax}%
\providecommand \BibitemShut  [1]{\csname bibitem#1\endcsname}%
\let\auto@bib@innerbib\@empty
\bibitem [{\citenamefont {{van Cittert}}(1934)}]{Cittert:34}%
  \BibitemOpen
  \bibfield  {author} {\bibinfo {author} {\bibfnamefont {P.}~\bibnamefont {{van
  Cittert}}},\ }\bibfield  {title} {\bibinfo {title} {Die wahrscheinliche
  schwingungsverteilung in einer von einer lichtquelle direkt oder mittels
  einer linse beleuchteten ebene},\ }\href
  {https://doi.org/https://doi.org/10.1016/S0031-8914(34)90026-4} {\bibfield
  {journal} {\bibinfo  {journal} {Physica}\ }\textbf {\bibinfo {volume} {1}},\
  \bibinfo {pages} {201} (\bibinfo {year} {1934})}\BibitemShut {NoStop}%
\bibitem [{\citenamefont {Zernike}(1938)}]{ZERNIKE:38}%
  \BibitemOpen
  \bibfield  {author} {\bibinfo {author} {\bibfnamefont {F.}~\bibnamefont
  {Zernike}},\ }\bibfield  {title} {\bibinfo {title} {The concept of degree of
  coherence and its application to optical problems},\ }\href
  {https://doi.org/https://doi.org/10.1016/S0031-8914(38)80203-2} {\bibfield
  {journal} {\bibinfo  {journal} {Physica}\ }\textbf {\bibinfo {volume} {5}},\
  \bibinfo {pages} {785} (\bibinfo {year} {1938})}\BibitemShut {NoStop}%
\bibitem [{\citenamefont {Born}\ and\ \citenamefont
  {Wolf}(2013)}]{born2013principles}%
  \BibitemOpen
  \bibfield  {author} {\bibinfo {author} {\bibfnamefont {M.}~\bibnamefont
  {Born}}\ and\ \bibinfo {author} {\bibfnamefont {E.}~\bibnamefont {Wolf}},\
  }\href@noop {} {\emph {\bibinfo {title} {Principles of optics:
  electromagnetic theory of propagation, interference and diffraction of
  light}}}\ (\bibinfo  {publisher} {Elsevier},\ \bibinfo {year}
  {2013})\BibitemShut {NoStop}%
\bibitem [{\citenamefont {Wolf}(1954)}]{Wolf:54}%
  \BibitemOpen
  \bibfield  {author} {\bibinfo {author} {\bibfnamefont {E.}~\bibnamefont
  {Wolf}},\ }\bibfield  {title} {\bibinfo {title} {Optics in terms of
  observable quantities},\ }\href {https://doi.org/10.1007/BF02781855}
  {\bibfield  {journal} {\bibinfo  {journal} {Il Nuovo Cimento (1943-1954)}\
  }\textbf {\bibinfo {volume} {12}},\ \bibinfo {pages} {884} (\bibinfo {year}
  {1954})}\BibitemShut {NoStop}%
\bibitem [{\citenamefont {Dorrer}(2004)}]{Dorrer:04}%
  \BibitemOpen
  \bibfield  {author} {\bibinfo {author} {\bibfnamefont {C.}~\bibnamefont
  {Dorrer}},\ }\bibfield  {title} {\bibinfo {title} {Temporal van
  {C}ittert-{Z}ernike theorem and its application to the measurement of
  chromatic dispersion},\ }\href {https://doi.org/10.1364/JOSAB.21.001417}
  {\bibfield  {journal} {\bibinfo  {journal} {J. Opt. Soc. Am. B}\ }\textbf
  {\bibinfo {volume} {21}},\ \bibinfo {pages} {1417} (\bibinfo {year}
  {2004})}\BibitemShut {NoStop}%
\bibitem [{\citenamefont {Gori}\ \emph {et~al.}(2000)\citenamefont {Gori},
  \citenamefont {Santarsiero}, \citenamefont {Borghi},\ and\ \citenamefont
  {Piquero}}]{gori2000use}%
  \BibitemOpen
  \bibfield  {author} {\bibinfo {author} {\bibfnamefont {F.}~\bibnamefont
  {Gori}}, \bibinfo {author} {\bibfnamefont {M.}~\bibnamefont {Santarsiero}},
  \bibinfo {author} {\bibfnamefont {R.}~\bibnamefont {Borghi}},\ and\ \bibinfo
  {author} {\bibfnamefont {G.}~\bibnamefont {Piquero}},\ }\bibfield  {title}
  {\bibinfo {title} {Use of the van {C}ittert--{Z}ernike theorem for partially
  polarized sources},\ }\href {https://doi.org/10.1364/OL.25.001291} {\bibfield
   {journal} {\bibinfo  {journal} {Opt. Lett.}\ }\textbf {\bibinfo {volume}
  {25}},\ \bibinfo {pages} {1291} (\bibinfo {year} {2000})}\BibitemShut
  {NoStop}%
\bibitem [{\citenamefont {Cai}\ \emph {et~al.}(2020)\citenamefont {Cai},
  \citenamefont {Zhang},\ and\ \citenamefont {Gbur}}]{cai2020}%
  \BibitemOpen
  \bibfield  {author} {\bibinfo {author} {\bibfnamefont {Y.}~\bibnamefont
  {Cai}}, \bibinfo {author} {\bibfnamefont {Y.}~\bibnamefont {Zhang}},\ and\
  \bibinfo {author} {\bibfnamefont {G.}~\bibnamefont {Gbur}},\ }\bibfield
  {title} {\bibinfo {title} {Partially coherent vortex beams of arbitrary
  radial order and a van cittert–zernike theorem for vortices},\ }\href
  {https://doi.org/10.1103/PhysRevA.101.043812} {\bibfield  {journal} {\bibinfo
   {journal} {Phys. Rev. A}\ }\textbf {\bibinfo {volume} {101}},\ \bibinfo
  {pages} {043812} (\bibinfo {year} {2020})}\BibitemShut {NoStop}%
\bibitem [{\citenamefont {Cai}\ \emph {et~al.}(2012)\citenamefont {Cai},
  \citenamefont {Dong},\ and\ \citenamefont {Hoenders}}]{Cai2012}%
  \BibitemOpen
  \bibfield  {author} {\bibinfo {author} {\bibfnamefont {Y.}~\bibnamefont
  {Cai}}, \bibinfo {author} {\bibfnamefont {Y.}~\bibnamefont {Dong}},\ and\
  \bibinfo {author} {\bibfnamefont {B.}~\bibnamefont {Hoenders}},\ }\bibfield
  {title} {\bibinfo {title} {Interdependence between the temporal and spatial
  longitudinal and transverse degrees of partial coherence and a generalization
  of the van cittert-zernike theorem},\ }\href
  {https://doi.org/10.1364/JOSAA.29.002542} {\bibfield  {journal} {\bibinfo
  {journal} {J. Opt. Soc. Amer. A}\ }\textbf {\bibinfo {volume} {29}},\
  \bibinfo {pages} {2542} (\bibinfo {year} {2012})}\BibitemShut {NoStop}%
\bibitem [{\citenamefont {Carozzi}\ and\ \citenamefont
  {Woan}(2009)}]{Carozzi:09}%
  \BibitemOpen
  \bibfield  {author} {\bibinfo {author} {\bibfnamefont {T.~D.}\ \bibnamefont
  {Carozzi}}\ and\ \bibinfo {author} {\bibfnamefont {G.}~\bibnamefont {Woan}},\
  }\bibfield  {title} {\bibinfo {title} {{A generalized measurement equation
  and van Cittert-Zernike theorem for wide-field radio astronomical
  interferometry}},\ }\href {https://doi.org/10.1111/j.1365-2966.2009.14642.x}
  {\bibfield  {journal} {\bibinfo  {journal} {Mon. Not. R. Astron Soc.}\
  }\textbf {\bibinfo {volume} {395}},\ \bibinfo {pages} {1558} (\bibinfo {year}
  {2009})}\BibitemShut {NoStop}%
\bibitem [{\citenamefont {Batarseh}\ \emph {et~al.}(2018)\citenamefont
  {Batarseh}, \citenamefont {Sukhov}, \citenamefont {Shen}, \citenamefont
  {Gemar}, \citenamefont {Rezvani},\ and\ \citenamefont
  {Dogariu}}]{Batarseh:18}%
  \BibitemOpen
  \bibfield  {author} {\bibinfo {author} {\bibfnamefont {M.}~\bibnamefont
  {Batarseh}}, \bibinfo {author} {\bibfnamefont {S.}~\bibnamefont {Sukhov}},
  \bibinfo {author} {\bibfnamefont {Z.}~\bibnamefont {Shen}}, \bibinfo {author}
  {\bibfnamefont {H.}~\bibnamefont {Gemar}}, \bibinfo {author} {\bibfnamefont
  {R.}~\bibnamefont {Rezvani}},\ and\ \bibinfo {author} {\bibfnamefont
  {A.}~\bibnamefont {Dogariu}},\ }\bibfield  {title} {\bibinfo {title} {Passive
  sensing around the corner using spatial coherence},\ }\href
  {https://doi.org/10.1038/s41467-018-05985-w} {\bibfield  {journal} {\bibinfo
  {journal} {Nat. Commun.}\ }\textbf {\bibinfo {volume} {9}},\ \bibinfo {pages}
  {3629} (\bibinfo {year} {2018})}\BibitemShut {NoStop}%
\bibitem [{\citenamefont {Barakat}(2000)}]{Barakat:2000}%
  \BibitemOpen
  \bibfield  {author} {\bibinfo {author} {\bibfnamefont {R.}~\bibnamefont
  {Barakat}},\ }\bibfield  {title} {\bibinfo {title} {Imaging via the van
  cittert zernike theorem using triple-correlations},\ }\href
  {https://doi.org/10.1080/09500340008231411} {\bibfield  {journal} {\bibinfo
  {journal} {J. Mod. Opt.}\ }\textbf {\bibinfo {volume} {47}},\ \bibinfo
  {pages} {1607} (\bibinfo {year} {2000})}\BibitemShut {NoStop}%
\bibitem [{\citenamefont {Barbosa}(1996)}]{PhysRevA.54.4473}%
  \BibitemOpen
  \bibfield  {author} {\bibinfo {author} {\bibfnamefont {G.~A.}\ \bibnamefont
  {Barbosa}},\ }\bibfield  {title} {\bibinfo {title} {Quantum images in
  double-slit experiments with spontaneous down-conversion light},\ }\href
  {https://doi.org/10.1103/PhysRevA.54.4473} {\bibfield  {journal} {\bibinfo
  {journal} {Phys. Rev. A}\ }\textbf {\bibinfo {volume} {54}},\ \bibinfo
  {pages} {4473} (\bibinfo {year} {1996})}\BibitemShut {NoStop}%
\bibitem [{\citenamefont {Saleh}\ \emph {et~al.}(2005)\citenamefont {Saleh},
  \citenamefont {Teich},\ and\ \citenamefont {Sergienko}}]{Saleh05PRL}%
  \BibitemOpen
  \bibfield  {author} {\bibinfo {author} {\bibfnamefont {B.~E.~A.}\
  \bibnamefont {Saleh}}, \bibinfo {author} {\bibfnamefont {M.~C.}\ \bibnamefont
  {Teich}},\ and\ \bibinfo {author} {\bibfnamefont {A.~V.}\ \bibnamefont
  {Sergienko}},\ }\bibfield  {title} {\bibinfo {title} {Wolf equations for
  two-photon light},\ }\href {https://doi.org/10.1103/PhysRevLett.94.223601}
  {\bibfield  {journal} {\bibinfo  {journal} {Phys. Rev. Lett.}\ }\textbf
  {\bibinfo {volume} {94}},\ \bibinfo {pages} {223601} (\bibinfo {year}
  {2005})}\BibitemShut {NoStop}%
\bibitem [{\citenamefont {Fabre}\ \emph {et~al.}(2017)\citenamefont {Fabre},
  \citenamefont {Navarrete}, \citenamefont {Sarkadi},\ and\ \citenamefont
  {Barrachina}}]{Fabre_2017}%
  \BibitemOpen
  \bibfield  {author} {\bibinfo {author} {\bibfnamefont {I.}~\bibnamefont
  {Fabre}}, \bibinfo {author} {\bibfnamefont {F.}~\bibnamefont {Navarrete}},
  \bibinfo {author} {\bibfnamefont {L.}~\bibnamefont {Sarkadi}},\ and\ \bibinfo
  {author} {\bibfnamefont {R.~O.}\ \bibnamefont {Barrachina}},\ }\bibfield
  {title} {\bibinfo {title} {Free evolution of an incoherent mixture of states:
  a quantum mechanical approach to the van cittert{\textendash}zernike
  theorem},\ }\href {https://doi.org/10.1088/1361-6404/aa8e74} {\bibfield
  {journal} {\bibinfo  {journal} {Eur. J. Phys.}\ }\textbf {\bibinfo {volume}
  {39}},\ \bibinfo {pages} {015401} (\bibinfo {year} {2017})}\BibitemShut
  {NoStop}%
\bibitem [{\citenamefont {Howard}\ \emph {et~al.}(2019)\citenamefont {Howard},
  \citenamefont {Gillett}, \citenamefont {Pearce}, \citenamefont {Abrahao},
  \citenamefont {Weinhold}, \citenamefont {Kok},\ and\ \citenamefont
  {White}}]{Howard2019}%
  \BibitemOpen
  \bibfield  {author} {\bibinfo {author} {\bibfnamefont {L.~A.}\ \bibnamefont
  {Howard}}, \bibinfo {author} {\bibfnamefont {G.~G.}\ \bibnamefont {Gillett}},
  \bibinfo {author} {\bibfnamefont {M.~E.}\ \bibnamefont {Pearce}}, \bibinfo
  {author} {\bibfnamefont {R.~A.}\ \bibnamefont {Abrahao}}, \bibinfo {author}
  {\bibfnamefont {T.~J.}\ \bibnamefont {Weinhold}}, \bibinfo {author}
  {\bibfnamefont {P.}~\bibnamefont {Kok}},\ and\ \bibinfo {author}
  {\bibfnamefont {A.~G.}\ \bibnamefont {White}},\ }\bibfield  {title} {\bibinfo
  {title} {Optimal imaging of remote bodies using quantum detectors},\ }\href
  {https://doi.org/10.1103/PhysRevLett.123.143604} {\bibfield  {journal}
  {\bibinfo  {journal} {Phys. Rev. Lett.}\ }\textbf {\bibinfo {volume} {123}},\
  \bibinfo {pages} {143604} (\bibinfo {year} {2019})}\BibitemShut {NoStop}%
\bibitem [{\citenamefont {Barrachina}\ \emph {et~al.}(2020)\citenamefont
  {Barrachina}, \citenamefont {Navarrete},\ and\ \citenamefont
  {Ciappina}}]{Barrachina2020}%
  \BibitemOpen
  \bibfield  {author} {\bibinfo {author} {\bibfnamefont {R.~O.}\ \bibnamefont
  {Barrachina}}, \bibinfo {author} {\bibfnamefont {F.}~\bibnamefont
  {Navarrete}},\ and\ \bibinfo {author} {\bibfnamefont {M.~F.}\ \bibnamefont
  {Ciappina}},\ }\bibfield  {title} {\bibinfo {title} {Quantum coherence
  enfeebled by classical uncertainties},\ }\href
  {https://doi.org/10.1103/PhysRevResearch.2.043353} {\bibfield  {journal}
  {\bibinfo  {journal} {Phys. Rev. Research}\ }\textbf {\bibinfo {volume}
  {2}},\ \bibinfo {pages} {043353} (\bibinfo {year} {2020})}\BibitemShut
  {NoStop}%
\bibitem [{\citenamefont {Khabiboulline}\ \emph {et~al.}(2019)\citenamefont
  {Khabiboulline}, \citenamefont {Borregaard}, \citenamefont {De~Greve},\ and\
  \citenamefont {Lukin}}]{PhysRevLett.123.070504}%
  \BibitemOpen
  \bibfield  {author} {\bibinfo {author} {\bibfnamefont {E.~T.}\ \bibnamefont
  {Khabiboulline}}, \bibinfo {author} {\bibfnamefont {J.}~\bibnamefont
  {Borregaard}}, \bibinfo {author} {\bibfnamefont {K.}~\bibnamefont
  {De~Greve}},\ and\ \bibinfo {author} {\bibfnamefont {M.~D.}\ \bibnamefont
  {Lukin}},\ }\bibfield  {title} {\bibinfo {title} {Optical interferometry with
  quantum networks},\ }\href {https://doi.org/10.1103/PhysRevLett.123.070504}
  {\bibfield  {journal} {\bibinfo  {journal} {Phys. Rev. Lett.}\ }\textbf
  {\bibinfo {volume} {123}},\ \bibinfo {pages} {070504} (\bibinfo {year}
  {2019})}\BibitemShut {NoStop}%
\bibitem [{\citenamefont {Reichert}\ \emph {et~al.}(2017)\citenamefont
  {Reichert}, \citenamefont {Sun},\ and\ \citenamefont
  {Fleischer}}]{PhysRevA.95.063836}%
  \BibitemOpen
  \bibfield  {author} {\bibinfo {author} {\bibfnamefont {M.}~\bibnamefont
  {Reichert}}, \bibinfo {author} {\bibfnamefont {X.}~\bibnamefont {Sun}},\ and\
  \bibinfo {author} {\bibfnamefont {J.~W.}\ \bibnamefont {Fleischer}},\
  }\bibfield  {title} {\bibinfo {title} {Quality of spatial entanglement
  propagation},\ }\href {https://doi.org/10.1103/PhysRevA.95.063836} {\bibfield
   {journal} {\bibinfo  {journal} {Phys. Rev. A}\ }\textbf {\bibinfo {volume}
  {95}},\ \bibinfo {pages} {063836} (\bibinfo {year} {2017})}\BibitemShut
  {NoStop}%
\bibitem [{\citenamefont {Defienne}\ and\ \citenamefont
  {Gigan}(2019)}]{PhysRevA.99.053831}%
  \BibitemOpen
  \bibfield  {author} {\bibinfo {author} {\bibfnamefont {H.}~\bibnamefont
  {Defienne}}\ and\ \bibinfo {author} {\bibfnamefont {S.}~\bibnamefont
  {Gigan}},\ }\bibfield  {title} {\bibinfo {title} {Spatially entangled
  photon-pair generation using a partial spatially coherent pump beam},\ }\href
  {https://doi.org/10.1103/PhysRevA.99.053831} {\bibfield  {journal} {\bibinfo
  {journal} {Phys. Rev. A}\ }\textbf {\bibinfo {volume} {99}},\ \bibinfo
  {pages} {053831} (\bibinfo {year} {2019})}\BibitemShut {NoStop}%
\bibitem [{\citenamefont {Qian}\ \emph {et~al.}(2018)\citenamefont {Qian},
  \citenamefont {Vamivakas},\ and\ \citenamefont {Eberly}}]{Qian:18}%
  \BibitemOpen
  \bibfield  {author} {\bibinfo {author} {\bibfnamefont {X.-F.}\ \bibnamefont
  {Qian}}, \bibinfo {author} {\bibfnamefont {A.~N.}\ \bibnamefont
  {Vamivakas}},\ and\ \bibinfo {author} {\bibfnamefont {J.~H.}\ \bibnamefont
  {Eberly}},\ }\bibfield  {title} {\bibinfo {title} {Entanglement limits
  duality and vice versa},\ }\href {https://doi.org/10.1364/OPTICA.5.000942}
  {\bibfield  {journal} {\bibinfo  {journal} {Optica}\ }\textbf {\bibinfo
  {volume} {5}},\ \bibinfo {pages} {942} (\bibinfo {year} {2018})}\BibitemShut
  {NoStop}%
\bibitem [{\citenamefont {Eberly}\ \emph {et~al.}(2016)\citenamefont {Eberly},
  \citenamefont {Qian}, \citenamefont {Qasimi}, \citenamefont {Ali},
  \citenamefont {Alonso}, \citenamefont {Guti{\'{e}}rrez-Cuevas}, \citenamefont
  {Little}, \citenamefont {Howell}, \citenamefont {Malhotra},\ and\
  \citenamefont {Vamivakas}}]{Eberly_2016}%
  \BibitemOpen
  \bibfield  {author} {\bibinfo {author} {\bibfnamefont {J.~H.}\ \bibnamefont
  {Eberly}}, \bibinfo {author} {\bibfnamefont {X.-F.}\ \bibnamefont {Qian}},
  \bibinfo {author} {\bibfnamefont {A.~A.}\ \bibnamefont {Qasimi}}, \bibinfo
  {author} {\bibfnamefont {H.}~\bibnamefont {Ali}}, \bibinfo {author}
  {\bibfnamefont {M.~A.}\ \bibnamefont {Alonso}}, \bibinfo {author}
  {\bibfnamefont {R.}~\bibnamefont {Guti{\'{e}}rrez-Cuevas}}, \bibinfo {author}
  {\bibfnamefont {B.~J.}\ \bibnamefont {Little}}, \bibinfo {author}
  {\bibfnamefont {J.~C.}\ \bibnamefont {Howell}}, \bibinfo {author}
  {\bibfnamefont {T.}~\bibnamefont {Malhotra}},\ and\ \bibinfo {author}
  {\bibfnamefont {A.~N.}\ \bibnamefont {Vamivakas}},\ }\bibfield  {title}
  {\bibinfo {title} {Quantum and classical optics{\textendash}emerging links},\
  }\href {https://doi.org/10.1088/0031-8949/91/6/063003} {\bibfield  {journal}
  {\bibinfo  {journal} {Phys. Scr.}\ }\textbf {\bibinfo {volume} {91}},\
  \bibinfo {pages} {063003} (\bibinfo {year} {2016})}\BibitemShut {NoStop}%
\bibitem [{\citenamefont {Bhusal}\ \emph {et~al.}(2022)\citenamefont {Bhusal},
  \citenamefont {Hong}, \citenamefont {Miller}, \citenamefont {Quiroz-Juárez},
  \citenamefont {León-Montiel}, \citenamefont {You},\ and\ \citenamefont
  {Magaña-Loaiza}}]{bhusal2021smart}%
  \BibitemOpen
  \bibfield  {author} {\bibinfo {author} {\bibfnamefont {N.}~\bibnamefont
  {Bhusal}}, \bibinfo {author} {\bibfnamefont {M.}~\bibnamefont {Hong}},
  \bibinfo {author} {\bibfnamefont {A.}~\bibnamefont {Miller}}, \bibinfo
  {author} {\bibfnamefont {M.~A.}\ \bibnamefont {Quiroz-Juárez}}, \bibinfo
  {author} {\bibfnamefont {R.~d.~J.}\ \bibnamefont {León-Montiel}}, \bibinfo
  {author} {\bibfnamefont {C.}~\bibnamefont {You}},\ and\ \bibinfo {author}
  {\bibfnamefont {O.~S.}\ \bibnamefont {Magaña-Loaiza}},\ }\bibfield  {title}
  {\bibinfo {title} {Smart quantum statistical imaging beyond the
  {Abbe}-{Rayleigh} criterion},\ }\href
  {https://doi.org/10.1038/s41534-022-00593-5} {\bibfield  {journal} {\bibinfo
  {journal} {NPJ Quantum Inf}\ }\textbf {\bibinfo {volume} {8}},\ \bibinfo
  {pages} {83} (\bibinfo {year} {2022})}\BibitemShut {NoStop}%
\bibitem [{\citenamefont {de~J.~Le{\'{o}}n-Montiel}\ \emph
  {et~al.}(2013)\citenamefont {de~J.~Le{\'{o}}n-Montiel}, \citenamefont
  {Svozil{\'{\i}}k}, \citenamefont {Salazar-Serrano},\ and\ \citenamefont
  {Torres}}]{de_J_Le_n_Montiel_2013}%
  \BibitemOpen
  \bibfield  {author} {\bibinfo {author} {\bibfnamefont {R.}~\bibnamefont
  {de~J.~Le{\'{o}}n-Montiel}}, \bibinfo {author} {\bibfnamefont
  {J.}~\bibnamefont {Svozil{\'{\i}}k}}, \bibinfo {author} {\bibfnamefont
  {L.~J.}\ \bibnamefont {Salazar-Serrano}},\ and\ \bibinfo {author}
  {\bibfnamefont {J.~P.}\ \bibnamefont {Torres}},\ }\bibfield  {title}
  {\bibinfo {title} {Role of the spectral shape of quantum correlations in
  two-photon virtual-state spectroscopy},\ }\href
  {https://doi.org/10.1088/1367-2630/15/5/053023} {\bibfield  {journal}
  {\bibinfo  {journal} {New. J. Phys.}\ }\textbf {\bibinfo {volume} {15}},\
  \bibinfo {pages} {053023} (\bibinfo {year} {2013})}\BibitemShut {NoStop}%
\bibitem [{\citenamefont {You}\ \emph {et~al.}(2021{\natexlab{a}})\citenamefont
  {You}, \citenamefont {Hong}, \citenamefont {Bierhorst}, \citenamefont {Lita},
  \citenamefont {Glancy}, \citenamefont {Kolthammer}, \citenamefont {Knill},
  \citenamefont {Nam}, \citenamefont {Mirin}, \citenamefont {Magaña-Loaiza},\
  and\ \citenamefont {Gerrits}}]{you2021scalable}%
  \BibitemOpen
  \bibfield  {author} {\bibinfo {author} {\bibfnamefont {C.}~\bibnamefont
  {You}}, \bibinfo {author} {\bibfnamefont {M.}~\bibnamefont {Hong}}, \bibinfo
  {author} {\bibfnamefont {P.}~\bibnamefont {Bierhorst}}, \bibinfo {author}
  {\bibfnamefont {A.~E.}\ \bibnamefont {Lita}}, \bibinfo {author}
  {\bibfnamefont {S.}~\bibnamefont {Glancy}}, \bibinfo {author} {\bibfnamefont
  {S.}~\bibnamefont {Kolthammer}}, \bibinfo {author} {\bibfnamefont
  {E.}~\bibnamefont {Knill}}, \bibinfo {author} {\bibfnamefont {S.~W.}\
  \bibnamefont {Nam}}, \bibinfo {author} {\bibfnamefont {R.~P.}\ \bibnamefont
  {Mirin}}, \bibinfo {author} {\bibfnamefont {O.~S.}\ \bibnamefont
  {Magaña-Loaiza}},\ and\ \bibinfo {author} {\bibfnamefont {T.}~\bibnamefont
  {Gerrits}},\ }\bibfield  {title} {\bibinfo {title} {Scalable multiphoton
  quantum metrology with neither pre- nor post-selected measurements},\ }\href
  {https://doi.org/10.1063/5.0063294} {\bibfield  {journal} {\bibinfo
  {journal} {Appl. Phys. Rev.}\ }\textbf {\bibinfo {volume} {8}},\ \bibinfo
  {pages} {041406} (\bibinfo {year} {2021}{\natexlab{a}})}\BibitemShut
  {NoStop}%
\bibitem [{\citenamefont {O'Brien}\ \emph {et~al.}(2009)\citenamefont
  {O'Brien}, \citenamefont {Furusawa},\ and\ \citenamefont
  {Vučković}}]{obrien_photonic_2009}%
  \BibitemOpen
  \bibfield  {author} {\bibinfo {author} {\bibfnamefont {J.~L.}\ \bibnamefont
  {O'Brien}}, \bibinfo {author} {\bibfnamefont {A.}~\bibnamefont {Furusawa}},\
  and\ \bibinfo {author} {\bibfnamefont {J.}~\bibnamefont {Vučković}},\
  }\bibfield  {title} {\bibinfo {title} {Photonic quantum technologies},\
  }\href {https://doi.org/10.1038/nphoton.2009.229} {\bibfield  {journal}
  {\bibinfo  {journal} {Nat. Photonics}\ }\textbf {\bibinfo {volume} {3}},\
  \bibinfo {pages} {687} (\bibinfo {year} {2009})}\BibitemShut {NoStop}%
\bibitem [{\citenamefont {Wen}\ \emph {et~al.}(2013)\citenamefont {Wen},
  \citenamefont {Zhang},\ and\ \citenamefont {Xiao}}]{Wen:13}%
  \BibitemOpen
  \bibfield  {author} {\bibinfo {author} {\bibfnamefont {J.}~\bibnamefont
  {Wen}}, \bibinfo {author} {\bibfnamefont {Y.}~\bibnamefont {Zhang}},\ and\
  \bibinfo {author} {\bibfnamefont {M.}~\bibnamefont {Xiao}},\ }\bibfield
  {title} {\bibinfo {title} {The talbot effect: recent advances in classical
  optics, nonlinear optics, and quantum optics},\ }\href
  {https://doi.org/10.1364/AOP.5.000083} {\bibfield  {journal} {\bibinfo
  {journal} {Adv. Opt. Photon.}\ }\textbf {\bibinfo {volume} {5}},\ \bibinfo
  {pages} {83} (\bibinfo {year} {2013})}\BibitemShut {NoStop}%
\bibitem [{\citenamefont {Maga\~na Loaiza}\ \emph {et~al.}(2019)\citenamefont
  {Maga\~na Loaiza}, \citenamefont {de~J.~Le{\'{o}}n-Montiel}, \citenamefont
  {Perez-Leija}, \citenamefont {U’Ren}, \citenamefont {You}, \citenamefont
  {Busch}, \citenamefont {Lita}, \citenamefont {Nam}, \citenamefont {Mirin},\
  and\ \citenamefont {Gerrits}}]{magana2019multiphoton}%
  \BibitemOpen
  \bibfield  {author} {\bibinfo {author} {\bibfnamefont {O.~S.}\ \bibnamefont
  {Maga\~na Loaiza}}, \bibinfo {author} {\bibfnamefont {R.}~\bibnamefont
  {de~J.~Le{\'{o}}n-Montiel}}, \bibinfo {author} {\bibfnamefont
  {A.}~\bibnamefont {Perez-Leija}}, \bibinfo {author} {\bibfnamefont {A.~B.}\
  \bibnamefont {U’Ren}}, \bibinfo {author} {\bibfnamefont {C.}~\bibnamefont
  {You}}, \bibinfo {author} {\bibfnamefont {K.}~\bibnamefont {Busch}}, \bibinfo
  {author} {\bibfnamefont {A.~E.}\ \bibnamefont {Lita}}, \bibinfo {author}
  {\bibfnamefont {S.~W.}\ \bibnamefont {Nam}}, \bibinfo {author} {\bibfnamefont
  {R.~P.}\ \bibnamefont {Mirin}},\ and\ \bibinfo {author} {\bibfnamefont
  {T.}~\bibnamefont {Gerrits}},\ }\bibfield  {title} {\bibinfo {title}
  {Multiphoton quantum-state engineering using conditional measurements},\
  }\href {https://doi.org/10.1038/s41534-019-0195-2} {\bibfield  {journal}
  {\bibinfo  {journal} {npj Quantum Inf.}\ }\textbf {\bibinfo {volume} {5}},\
  \bibinfo {pages} {80} (\bibinfo {year} {2019})}\BibitemShut {NoStop}%
\bibitem [{\citenamefont {Dell’Anno}\ \emph {et~al.}(2006)\citenamefont
  {Dell’Anno}, \citenamefont {{De Siena}},\ and\ \citenamefont
  {Illuminati}}]{DELLANNO200653}%
  \BibitemOpen
  \bibfield  {author} {\bibinfo {author} {\bibfnamefont {F.}~\bibnamefont
  {Dell’Anno}}, \bibinfo {author} {\bibfnamefont {S.}~\bibnamefont {{De
  Siena}}},\ and\ \bibinfo {author} {\bibfnamefont {F.}~\bibnamefont
  {Illuminati}},\ }\bibfield  {title} {\bibinfo {title} {Multiphoton quantum
  optics and quantum state engineering},\ }\href
  {https://doi.org/https://doi.org/10.1016/j.physrep.2006.01.004} {\bibfield
  {journal} {\bibinfo  {journal} {Phys. Rep.}\ }\textbf {\bibinfo {volume}
  {428}},\ \bibinfo {pages} {53} (\bibinfo {year} {2006})}\BibitemShut
  {NoStop}%
\bibitem [{\citenamefont {Olsen}\ \emph {et~al.}(2002)\citenamefont {Olsen},
  \citenamefont {Plimak},\ and\ \citenamefont {Khoury}}]{OLSEN2002373}%
  \BibitemOpen
  \bibfield  {author} {\bibinfo {author} {\bibfnamefont {M.}~\bibnamefont
  {Olsen}}, \bibinfo {author} {\bibfnamefont {L.}~\bibnamefont {Plimak}},\ and\
  \bibinfo {author} {\bibfnamefont {A.}~\bibnamefont {Khoury}},\ }\bibfield
  {title} {\bibinfo {title} {Dynamical quantum statistical effects in optical
  parametric processes},\ }\href
  {https://doi.org/https://doi.org/10.1016/S0030-4018(01)01711-4} {\bibfield
  {journal} {\bibinfo  {journal} {Optics Communications}\ }\textbf {\bibinfo
  {volume} {201}},\ \bibinfo {pages} {373} (\bibinfo {year}
  {2002})}\BibitemShut {NoStop}%
\bibitem [{\citenamefont {Muñoz}\ \emph {et~al.}(2014)\citenamefont {Muñoz},
  \citenamefont {del Valle}, \citenamefont {Tudela}, \citenamefont {Müller},
  \citenamefont {Lichtmannecker}, \citenamefont {Kaniber}, \citenamefont
  {Tejedor}, \citenamefont {Finley},\ and\ \citenamefont
  {Laussy}}]{munoz_emitters_2014}%
  \BibitemOpen
  \bibfield  {author} {\bibinfo {author} {\bibfnamefont {C.~S.}\ \bibnamefont
  {Muñoz}}, \bibinfo {author} {\bibfnamefont {E.}~\bibnamefont {del Valle}},
  \bibinfo {author} {\bibfnamefont {A.~G.}\ \bibnamefont {Tudela}}, \bibinfo
  {author} {\bibfnamefont {K.}~\bibnamefont {Müller}}, \bibinfo {author}
  {\bibfnamefont {S.}~\bibnamefont {Lichtmannecker}}, \bibinfo {author}
  {\bibfnamefont {M.}~\bibnamefont {Kaniber}}, \bibinfo {author} {\bibfnamefont
  {C.}~\bibnamefont {Tejedor}}, \bibinfo {author} {\bibfnamefont {J.~J.}\
  \bibnamefont {Finley}},\ and\ \bibinfo {author} {\bibfnamefont {F.~P.}\
  \bibnamefont {Laussy}},\ }\bibfield  {title} {\bibinfo {title} {Emitters of
  {N}-photon bundles},\ }\href {https://doi.org/10.1038/nphoton.2014.114}
  {\bibfield  {journal} {\bibinfo  {journal} {Nat. Photonics}\ }\textbf
  {\bibinfo {volume} {8}},\ \bibinfo {pages} {550} (\bibinfo {year}
  {2014})}\BibitemShut {NoStop}%
\bibitem [{\citenamefont {Aspuru-Guzik}\ and\ \citenamefont
  {Walther}(2012)}]{aspuru-guzik_photonic_2012}%
  \BibitemOpen
  \bibfield  {author} {\bibinfo {author} {\bibfnamefont {A.}~\bibnamefont
  {Aspuru-Guzik}}\ and\ \bibinfo {author} {\bibfnamefont {P.}~\bibnamefont
  {Walther}},\ }\bibfield  {title} {\bibinfo {title} {Photonic quantum
  simulators},\ }\href {https://doi.org/10.1038/nphys2253} {\bibfield
  {journal} {\bibinfo  {journal} {Nat. Phys.}\ }\textbf {\bibinfo {volume}
  {8}},\ \bibinfo {pages} {285} (\bibinfo {year} {2012})}\BibitemShut {NoStop}%
\bibitem [{\citenamefont {You}\ \emph {et~al.}(2020{\natexlab{a}})\citenamefont
  {You}, \citenamefont {Nellikka}, \citenamefont {Leon},\ and\ \citenamefont
  {Magaña-Loaiza}}]{You2020plasmonics}%
  \BibitemOpen
  \bibfield  {author} {\bibinfo {author} {\bibfnamefont {C.}~\bibnamefont
  {You}}, \bibinfo {author} {\bibfnamefont {A.~C.}\ \bibnamefont {Nellikka}},
  \bibinfo {author} {\bibfnamefont {I.~D.}\ \bibnamefont {Leon}},\ and\
  \bibinfo {author} {\bibfnamefont {O.~S.}\ \bibnamefont {Magaña-Loaiza}},\
  }\bibfield  {title} {\bibinfo {title} {Multiparticle quantum plasmonics},\
  }\href {https://doi.org/doi:10.1515/nanoph-2019-0517} {\bibfield  {journal}
  {\bibinfo  {journal} {Nanophotonics}\ }\textbf {\bibinfo {volume} {9}},\
  \bibinfo {pages} {1243} (\bibinfo {year} {2020}{\natexlab{a}})}\BibitemShut
  {NoStop}%
\bibitem [{\citenamefont {Mandel}(1979)}]{Mandel:79}%
  \BibitemOpen
  \bibfield  {author} {\bibinfo {author} {\bibfnamefont {L.}~\bibnamefont
  {Mandel}},\ }\bibfield  {title} {\bibinfo {title} {Sub-poissonian photon
  statistics in resonance fluorescence},\ }\href
  {https://doi.org/10.1364/OL.4.000205} {\bibfield  {journal} {\bibinfo
  {journal} {Opt. Lett.}\ }\textbf {\bibinfo {volume} {4}},\ \bibinfo {pages}
  {205} (\bibinfo {year} {1979})}\BibitemShut {NoStop}%
\bibitem [{\citenamefont {You}\ \emph {et~al.}(2020{\natexlab{b}})\citenamefont
  {You}, \citenamefont {Quiroz-Ju{\'a}rez}, \citenamefont {Lambert},
  \citenamefont {Bhusal}, \citenamefont {Dong}, \citenamefont {Perez-Leija},
  \citenamefont {Javaid}, \citenamefont {de~J.~Le{\'{o}}n-Montiel},\ and\
  \citenamefont {Maga{\~n}a-Loaiza}}]{you2020identification}%
  \BibitemOpen
  \bibfield  {author} {\bibinfo {author} {\bibfnamefont {C.}~\bibnamefont
  {You}}, \bibinfo {author} {\bibfnamefont {M.~A.}\ \bibnamefont
  {Quiroz-Ju{\'a}rez}}, \bibinfo {author} {\bibfnamefont {A.}~\bibnamefont
  {Lambert}}, \bibinfo {author} {\bibfnamefont {N.}~\bibnamefont {Bhusal}},
  \bibinfo {author} {\bibfnamefont {C.}~\bibnamefont {Dong}}, \bibinfo {author}
  {\bibfnamefont {A.}~\bibnamefont {Perez-Leija}}, \bibinfo {author}
  {\bibfnamefont {A.}~\bibnamefont {Javaid}}, \bibinfo {author} {\bibfnamefont
  {R.}~\bibnamefont {de~J.~Le{\'{o}}n-Montiel}},\ and\ \bibinfo {author}
  {\bibfnamefont {O.~S.}\ \bibnamefont {Maga{\~n}a-Loaiza}},\ }\bibfield
  {title} {\bibinfo {title} {Identification of light sources using machine
  learning},\ }\href {https://doi.org/10.1063/1.5133846} {\bibfield  {journal}
  {\bibinfo  {journal} {Appl. Phys. Rev.}\ }\textbf {\bibinfo {volume} {7}},\
  \bibinfo {pages} {021404} (\bibinfo {year} {2020}{\natexlab{b}})}\BibitemShut
  {NoStop}%
\bibitem [{\citenamefont {Mandel}\ and\ \citenamefont
  {Wolf}(1995)}]{mandel1995optical}%
  \BibitemOpen
  \bibfield  {author} {\bibinfo {author} {\bibfnamefont {L.}~\bibnamefont
  {Mandel}}\ and\ \bibinfo {author} {\bibfnamefont {E.}~\bibnamefont {Wolf}},\
  }\href@noop {} {\emph {\bibinfo {title} {Optical coherence and quantum
  optics}}}\ (\bibinfo  {publisher} {Cambridge university press},\ \bibinfo
  {year} {1995})\BibitemShut {NoStop}%
\bibitem [{\citenamefont {Venkataraman}\ \emph {et~al.}(2013)\citenamefont
  {Venkataraman}, \citenamefont {Saha},\ and\ \citenamefont
  {Gaeta}}]{venkataraman_phase_2013}%
  \BibitemOpen
  \bibfield  {author} {\bibinfo {author} {\bibfnamefont {V.}~\bibnamefont
  {Venkataraman}}, \bibinfo {author} {\bibfnamefont {K.}~\bibnamefont {Saha}},\
  and\ \bibinfo {author} {\bibfnamefont {A.~L.}\ \bibnamefont {Gaeta}},\
  }\bibfield  {title} {\bibinfo {title} {Phase modulation at the few-photon
  level for weak-nonlinearity-based quantum computing},\ }\href
  {https://doi.org/10.1038/nphoton.2012.283} {\bibfield  {journal} {\bibinfo
  {journal} {Nature Photonics}\ }\textbf {\bibinfo {volume} {7}},\ \bibinfo
  {pages} {138} (\bibinfo {year} {2013})}\BibitemShut {NoStop}%
\bibitem [{\citenamefont {You}\ \emph {et~al.}(2021{\natexlab{b}})\citenamefont
  {You}, \citenamefont {Hong}, \citenamefont {Bhusal}, \citenamefont {Chen},
  \citenamefont {Quiroz-Ju{\'a}rez}, \citenamefont {Fabre}, \citenamefont
  {Mostafavi}, \citenamefont {Guo}, \citenamefont {De~Leon}, \citenamefont
  {de~J.~Le{\'{o}}n-Montiel},\ and\ \citenamefont
  {na~Loaiza}}]{you2021observation}%
  \BibitemOpen
  \bibfield  {author} {\bibinfo {author} {\bibfnamefont {C.}~\bibnamefont
  {You}}, \bibinfo {author} {\bibfnamefont {M.}~\bibnamefont {Hong}}, \bibinfo
  {author} {\bibfnamefont {N.}~\bibnamefont {Bhusal}}, \bibinfo {author}
  {\bibfnamefont {J.}~\bibnamefont {Chen}}, \bibinfo {author} {\bibfnamefont
  {M.~A.}\ \bibnamefont {Quiroz-Ju{\'a}rez}}, \bibinfo {author} {\bibfnamefont
  {J.}~\bibnamefont {Fabre}}, \bibinfo {author} {\bibfnamefont
  {F.}~\bibnamefont {Mostafavi}}, \bibinfo {author} {\bibfnamefont
  {J.}~\bibnamefont {Guo}}, \bibinfo {author} {\bibfnamefont {I.}~\bibnamefont
  {De~Leon}}, \bibinfo {author} {\bibfnamefont {R.}~\bibnamefont
  {de~J.~Le{\'{o}}n-Montiel}},\ and\ \bibinfo {author} {\bibfnamefont
  {O.~S.~M.}\ \bibnamefont {na~Loaiza}},\ }\bibfield  {title} {\bibinfo {title}
  {Observation of the modification of quantum statistics of plasmonic
  systems},\ }\href
  {https://doi.org/https://doi.org/10.1038/s41467-021-25489-4} {\bibfield
  {journal} {\bibinfo  {journal} {Nat. Commun.}\ }\textbf {\bibinfo {volume}
  {12}},\ \bibinfo {pages} {5161} (\bibinfo {year}
  {2021}{\natexlab{b}})}\BibitemShut {NoStop}%
\bibitem [{\citenamefont {Tame}(2021)}]{tame_mix_2021}%
  \BibitemOpen
  \bibfield  {author} {\bibinfo {author} {\bibfnamefont {M.}~\bibnamefont
  {Tame}},\ }\bibfield  {title} {\bibinfo {title} {Mix and match},\ }\href
  {https://doi.org/10.1038/s41567-021-01399-6} {\bibfield  {journal} {\bibinfo
  {journal} {Nature Physics}\ }\textbf {\bibinfo {volume} {17}},\ \bibinfo
  {pages} {1198} (\bibinfo {year} {2021})}\BibitemShut {NoStop}%
\bibitem [{\citenamefont {Knill}\ \emph {et~al.}(2001)\citenamefont {Knill},
  \citenamefont {Laflamme},\ and\ \citenamefont {Milburn}}]{knill2001}%
  \BibitemOpen
  \bibfield  {author} {\bibinfo {author} {\bibfnamefont {E.}~\bibnamefont
  {Knill}}, \bibinfo {author} {\bibfnamefont {R.}~\bibnamefont {Laflamme}},\
  and\ \bibinfo {author} {\bibfnamefont {G.~J.}\ \bibnamefont {Milburn}},\
  }\bibfield  {title} {\bibinfo {title} {A scheme for efficient quantum
  computation with linear optics},\ }\href {https://doi.org/10.1038/35051009}
  {\bibfield  {journal} {\bibinfo  {journal} {Nature}\ }\textbf {\bibinfo
  {volume} {409}},\ \bibinfo {pages} {46} (\bibinfo {year} {2001})}\BibitemShut
  {NoStop}%
\bibitem [{\citenamefont {Kok}\ \emph {et~al.}(2007)\citenamefont {Kok},
  \citenamefont {Munro}, \citenamefont {Nemoto}, \citenamefont {Ralph},
  \citenamefont {Dowling},\ and\ \citenamefont {Milburn}}]{RevModPhys.79.135}%
  \BibitemOpen
  \bibfield  {author} {\bibinfo {author} {\bibfnamefont {P.}~\bibnamefont
  {Kok}}, \bibinfo {author} {\bibfnamefont {W.~J.}\ \bibnamefont {Munro}},
  \bibinfo {author} {\bibfnamefont {K.}~\bibnamefont {Nemoto}}, \bibinfo
  {author} {\bibfnamefont {T.~C.}\ \bibnamefont {Ralph}}, \bibinfo {author}
  {\bibfnamefont {J.~P.}\ \bibnamefont {Dowling}},\ and\ \bibinfo {author}
  {\bibfnamefont {G.~J.}\ \bibnamefont {Milburn}},\ }\bibfield  {title}
  {\bibinfo {title} {Linear optical quantum computing with photonic qubits},\
  }\href {https://doi.org/10.1103/RevModPhys.79.135} {\bibfield  {journal}
  {\bibinfo  {journal} {Rev. Mod. Phys.}\ }\textbf {\bibinfo {volume} {79}},\
  \bibinfo {pages} {135} (\bibinfo {year} {2007})}\BibitemShut {NoStop}%
\bibitem [{\citenamefont {Yu}\ and\ \citenamefont
  {Eberly}(2009)}]{Yu_Sudden_2009}%
  \BibitemOpen
  \bibfield  {author} {\bibinfo {author} {\bibfnamefont {T.}~\bibnamefont
  {Yu}}\ and\ \bibinfo {author} {\bibfnamefont {J.~H.}\ \bibnamefont
  {Eberly}},\ }\bibfield  {title} {\bibinfo {title} {Sudden death of
  entanglement},\ }\href {https://doi.org/10.1126/science.1167343} {\bibfield
  {journal} {\bibinfo  {journal} {Science}\ }\textbf {\bibinfo {volume}
  {323}},\ \bibinfo {pages} {598} (\bibinfo {year} {2009})},\ \Eprint
  {https://arxiv.org/abs/https://www.science.org/doi/pdf/10.1126/science.1167343}
  {https://www.science.org/doi/pdf/10.1126/science.1167343} \BibitemShut
  {NoStop}%
\bibitem [{\citenamefont {Polino}\ \emph {et~al.}(2020)\citenamefont {Polino},
  \citenamefont {Valeri}, \citenamefont {Spagnolo},\ and\ \citenamefont
  {Sciarrino}}]{Polino_Photonic_2020}%
  \BibitemOpen
  \bibfield  {author} {\bibinfo {author} {\bibfnamefont {E.}~\bibnamefont
  {Polino}}, \bibinfo {author} {\bibfnamefont {M.}~\bibnamefont {Valeri}},
  \bibinfo {author} {\bibfnamefont {N.}~\bibnamefont {Spagnolo}},\ and\
  \bibinfo {author} {\bibfnamefont {F.}~\bibnamefont {Sciarrino}},\ }\bibfield
  {title} {\bibinfo {title} {Photonic quantum metrology},\ }\href
  {https://doi.org/10.1116/5.0007577} {\bibfield  {journal} {\bibinfo
  {journal} {AVS Quantum Science}\ }\textbf {\bibinfo {volume} {2}},\ \bibinfo
  {pages} {024703} (\bibinfo {year} {2020})},\ \Eprint
  {https://arxiv.org/abs/https://doi.org/10.1116/5.0007577}
  {https://doi.org/10.1116/5.0007577} \BibitemShut {NoStop}%
\bibitem [{\citenamefont {Kondakci}\ \emph {et~al.}(2015)\citenamefont
  {Kondakci}, \citenamefont {Abouraddy},\ and\ \citenamefont
  {Saleh}}]{kondakci_photonic_2015}%
  \BibitemOpen
  \bibfield  {author} {\bibinfo {author} {\bibfnamefont {H.~E.}\ \bibnamefont
  {Kondakci}}, \bibinfo {author} {\bibfnamefont {A.~F.}\ \bibnamefont
  {Abouraddy}},\ and\ \bibinfo {author} {\bibfnamefont {B.~E.~A.}\ \bibnamefont
  {Saleh}},\ }\bibfield  {title} {\bibinfo {title} {A photonic thermalization
  gap in disordered lattices},\ }\href {https://doi.org/10.1038/nphys3482}
  {\bibfield  {journal} {\bibinfo  {journal} {Nat. Phys.}\ }\textbf {\bibinfo
  {volume} {11}},\ \bibinfo {pages} {930} (\bibinfo {year} {2015})}\BibitemShut
  {NoStop}%
\bibitem [{\citenamefont {Magaña-Loaiza}\ \emph {et~al.}(2016)\citenamefont
  {Magaña-Loaiza}, \citenamefont {Mirhosseini}, \citenamefont {Cross},
  \citenamefont {Rafsanjani},\ and\ \citenamefont {Boyd}}]{magana-loaiza-2016}%
  \BibitemOpen
  \bibfield  {author} {\bibinfo {author} {\bibfnamefont {O.~S.}\ \bibnamefont
  {Magaña-Loaiza}}, \bibinfo {author} {\bibfnamefont {M.}~\bibnamefont
  {Mirhosseini}}, \bibinfo {author} {\bibfnamefont {R.~M.}\ \bibnamefont
  {Cross}}, \bibinfo {author} {\bibfnamefont {S.~M.~H.}\ \bibnamefont
  {Rafsanjani}},\ and\ \bibinfo {author} {\bibfnamefont {R.~W.}\ \bibnamefont
  {Boyd}},\ }\bibfield  {title} {\bibinfo {title} {Hanbury brown and twiss
  interferometry with twisted light},\ }\href
  {https://doi.org/10.1126/sciadv.1501143} {\bibfield  {journal} {\bibinfo
  {journal} {Sci. Adv.}\ }\textbf {\bibinfo {volume} {2}},\ \bibinfo {pages}
  {e1501143} (\bibinfo {year} {2016})}\BibitemShut {NoStop}%
\bibitem [{\citenamefont {Liu}\ and\ \citenamefont {Shih}(2009)}]{Shih2009PRA}%
  \BibitemOpen
  \bibfield  {author} {\bibinfo {author} {\bibfnamefont {J.}~\bibnamefont
  {Liu}}\ and\ \bibinfo {author} {\bibfnamefont {Y.}~\bibnamefont {Shih}},\
  }\bibfield  {title} {\bibinfo {title} {$n\text{th}$-order coherence of
  thermal light},\ }\href {https://doi.org/10.1103/PhysRevA.79.023819}
  {\bibfield  {journal} {\bibinfo  {journal} {Phys. Rev. A}\ }\textbf {\bibinfo
  {volume} {79}},\ \bibinfo {pages} {023819} (\bibinfo {year}
  {2009})}\BibitemShut {NoStop}%
\bibitem [{\citenamefont {Agafonov}\ \emph {et~al.}(2008)\citenamefont
  {Agafonov}, \citenamefont {Chekhova}, \citenamefont {Iskhakov},\ and\
  \citenamefont {Penin}}]{Chekhova08PRA}%
  \BibitemOpen
  \bibfield  {author} {\bibinfo {author} {\bibfnamefont {I.~N.}\ \bibnamefont
  {Agafonov}}, \bibinfo {author} {\bibfnamefont {M.~V.}\ \bibnamefont
  {Chekhova}}, \bibinfo {author} {\bibfnamefont {T.~S.}\ \bibnamefont
  {Iskhakov}},\ and\ \bibinfo {author} {\bibfnamefont {A.~N.}\ \bibnamefont
  {Penin}},\ }\bibfield  {title} {\bibinfo {title} {High-visibility multiphoton
  interference of hanbury brown--twiss type for classical light},\ }\href
  {https://doi.org/10.1103/PhysRevA.77.053801} {\bibfield  {journal} {\bibinfo
  {journal} {Phys. Rev. A}\ }\textbf {\bibinfo {volume} {77}},\ \bibinfo
  {pages} {053801} (\bibinfo {year} {2008})}\BibitemShut {NoStop}%
\bibitem [{\citenamefont {S{\"o}derholm}\ \emph {et~al.}(2001)\citenamefont
  {S{\"o}derholm}, \citenamefont {Bj{\"o}rk},\ and\ \citenamefont
  {Trifonov}}]{soderholm2001unpolarized}%
  \BibitemOpen
  \bibfield  {author} {\bibinfo {author} {\bibfnamefont {J.}~\bibnamefont
  {S{\"o}derholm}}, \bibinfo {author} {\bibfnamefont {G.}~\bibnamefont
  {Bj{\"o}rk}},\ and\ \bibinfo {author} {\bibfnamefont {A.}~\bibnamefont
  {Trifonov}},\ }\bibfield  {title} {\bibinfo {title} {Unpolarized light in
  quantum optics},\ }\href {https://doi.org/10.1134/1.14126677} {\bibfield
  {journal} {\bibinfo  {journal} {Opt. Spectrosc.}\ }\textbf {\bibinfo {volume}
  {91}},\ \bibinfo {pages} {532} (\bibinfo {year} {2001})}\BibitemShut
  {NoStop}%
\bibitem [{\citenamefont {Wei}\ \emph {et~al.}(2005)\citenamefont {Wei},
  \citenamefont {Altepeter}, \citenamefont {Branning}, \citenamefont
  {Goldbart}, \citenamefont {James}, \citenamefont {Jeffrey}, \citenamefont
  {Kwiat}, \citenamefont {Mukhopadhyay},\ and\ \citenamefont
  {Peters}}]{wei2005synthesizing}%
  \BibitemOpen
  \bibfield  {author} {\bibinfo {author} {\bibfnamefont {T.-C.}\ \bibnamefont
  {Wei}}, \bibinfo {author} {\bibfnamefont {J.~B.}\ \bibnamefont {Altepeter}},
  \bibinfo {author} {\bibfnamefont {D.}~\bibnamefont {Branning}}, \bibinfo
  {author} {\bibfnamefont {P.~M.}\ \bibnamefont {Goldbart}}, \bibinfo {author}
  {\bibfnamefont {D.~F.~V.}\ \bibnamefont {James}}, \bibinfo {author}
  {\bibfnamefont {E.}~\bibnamefont {Jeffrey}}, \bibinfo {author} {\bibfnamefont
  {P.~G.}\ \bibnamefont {Kwiat}}, \bibinfo {author} {\bibfnamefont
  {S.}~\bibnamefont {Mukhopadhyay}},\ and\ \bibinfo {author} {\bibfnamefont
  {N.~A.}\ \bibnamefont {Peters}},\ }\bibfield  {title} {\bibinfo {title}
  {Synthesizing arbitrary two-photon polarization mixed states},\ }\href
  {https://doi.org/10.1103/PhysRevA.71.032329} {\bibfield  {journal} {\bibinfo
  {journal} {Phys. Rev. A}\ }\textbf {\bibinfo {volume} {71}},\ \bibinfo
  {pages} {032329} (\bibinfo {year} {2005})}\BibitemShut {NoStop}%
\bibitem [{\citenamefont {Gureyev}\ \emph {et~al.}(2017)\citenamefont
  {Gureyev}, \citenamefont {Kozlov}, \citenamefont {Paganin}, \citenamefont
  {Nesterets}, \citenamefont {Hoog},\ and\ \citenamefont
  {Quiney}}]{gureyev2017van}%
  \BibitemOpen
  \bibfield  {author} {\bibinfo {author} {\bibfnamefont {T.~E.}\ \bibnamefont
  {Gureyev}}, \bibinfo {author} {\bibfnamefont {A.}~\bibnamefont {Kozlov}},
  \bibinfo {author} {\bibfnamefont {D.~M.}\ \bibnamefont {Paganin}}, \bibinfo
  {author} {\bibfnamefont {Y.~I.}\ \bibnamefont {Nesterets}}, \bibinfo {author}
  {\bibfnamefont {F.~D.}\ \bibnamefont {Hoog}},\ and\ \bibinfo {author}
  {\bibfnamefont {H.~M.}\ \bibnamefont {Quiney}},\ }\bibfield  {title}
  {\bibinfo {title} {On the van {C}ittert--{Z}ernike theorem for intensity
  correlations and its applications},\ }\href
  {https://doi.org/10.1364/JOSAA.34.001577} {\bibfield  {journal} {\bibinfo
  {journal} {J. Opt. Soc. Am. A}\ }\textbf {\bibinfo {volume} {34}},\ \bibinfo
  {pages} {1577} (\bibinfo {year} {2017})}\BibitemShut {NoStop}%
\bibitem [{\citenamefont {Perez-Leija}\ \emph {et~al.}(2017)\citenamefont
  {Perez-Leija}, \citenamefont {de~J.~Le{\'{o}}n-Montiel}, \citenamefont
  {Sperling}, \citenamefont {Moya-Cessa}, \citenamefont {Szameit},\ and\
  \citenamefont {Busch}}]{perez2017two}%
  \BibitemOpen
  \bibfield  {author} {\bibinfo {author} {\bibfnamefont {A.}~\bibnamefont
  {Perez-Leija}}, \bibinfo {author} {\bibfnamefont {R.}~\bibnamefont
  {de~J.~Le{\'{o}}n-Montiel}}, \bibinfo {author} {\bibfnamefont
  {J.}~\bibnamefont {Sperling}}, \bibinfo {author} {\bibfnamefont
  {H.}~\bibnamefont {Moya-Cessa}}, \bibinfo {author} {\bibfnamefont
  {A.}~\bibnamefont {Szameit}},\ and\ \bibinfo {author} {\bibfnamefont
  {K.}~\bibnamefont {Busch}},\ }\bibfield  {title} {\bibinfo {title}
  {Two-particle four-point correlations in dynamically disordered tight-binding
  networks},\ }\href {https://doi.org/10.1088/1361-6455/aa9aa1} {\bibfield
  {journal} {\bibinfo  {journal} {J. Phys. B: At. Mol. Opt. Phys.}\ }\textbf
  {\bibinfo {volume} {51}},\ \bibinfo {pages} {024002} (\bibinfo {year}
  {2017})}\BibitemShut {NoStop}%
\bibitem [{\citenamefont {Gerry}\ \emph {et~al.}(2005)\citenamefont {Gerry},
  \citenamefont {Knight},\ and\ \citenamefont
  {Knight}}]{gerry2005introductory}%
  \BibitemOpen
  \bibfield  {author} {\bibinfo {author} {\bibfnamefont {C.}~\bibnamefont
  {Gerry}}, \bibinfo {author} {\bibfnamefont {P.}~\bibnamefont {Knight}},\ and\
  \bibinfo {author} {\bibfnamefont {P.~L.}\ \bibnamefont {Knight}},\
  }\href@noop {} {\emph {\bibinfo {title} {Introductory quantum optics}}}\
  (\bibinfo  {publisher} {Cambridge university press},\ \bibinfo {year}
  {2005})\BibitemShut {NoStop}%
\bibitem [{\citenamefont {Cramer}\ \emph {et~al.}(2010)\citenamefont {Cramer},
  \citenamefont {Plenio}, \citenamefont {Flammia}, \citenamefont {Somma},
  \citenamefont {Gross}, \citenamefont {Bartlett}, \citenamefont
  {Landon-Cardinal}, \citenamefont {Poulin},\ and\ \citenamefont
  {Liu}}]{cramer2010efficient}%
  \BibitemOpen
  \bibfield  {author} {\bibinfo {author} {\bibfnamefont {M.}~\bibnamefont
  {Cramer}}, \bibinfo {author} {\bibfnamefont {M.~B.}\ \bibnamefont {Plenio}},
  \bibinfo {author} {\bibfnamefont {S.~T.}\ \bibnamefont {Flammia}}, \bibinfo
  {author} {\bibfnamefont {R.}~\bibnamefont {Somma}}, \bibinfo {author}
  {\bibfnamefont {D.}~\bibnamefont {Gross}}, \bibinfo {author} {\bibfnamefont
  {S.~D.}\ \bibnamefont {Bartlett}}, \bibinfo {author} {\bibfnamefont
  {O.}~\bibnamefont {Landon-Cardinal}}, \bibinfo {author} {\bibfnamefont
  {D.}~\bibnamefont {Poulin}},\ and\ \bibinfo {author} {\bibfnamefont {Y.-K.}\
  \bibnamefont {Liu}},\ }\bibfield  {title} {\bibinfo {title} {Efficient
  quantum state tomography},\ }\href {https://doi.org/10.1038/ncomms1147}
  {\bibfield  {journal} {\bibinfo  {journal} {Nat. Commun.}\ }\textbf {\bibinfo
  {volume} {1}},\ \bibinfo {pages} {149} (\bibinfo {year} {2010})}\BibitemShut
  {NoStop}%
\bibitem [{\citenamefont {Aaronson}\ and\ \citenamefont
  {Arkhipov}(2011)}]{10.1145/1993636.1993682}%
  \BibitemOpen
  \bibfield  {author} {\bibinfo {author} {\bibfnamefont {S.}~\bibnamefont
  {Aaronson}}\ and\ \bibinfo {author} {\bibfnamefont {A.}~\bibnamefont
  {Arkhipov}},\ }\bibfield  {title} {\bibinfo {title} {The computational
  complexity of linear optics},\ }in\ \href
  {https://doi.org/10.1145/1993636.1993682} {\emph {\bibinfo {booktitle}
  {Proceedings of the Forty-Third Annual ACM Symposium on Theory of
  Computing}}},\ \bibinfo {series and number} {STOC '11}\ (\bibinfo
  {publisher} {Association for Computing Machinery},\ \bibinfo {address} {New
  York, NY, USA},\ \bibinfo {year} {2011})\ p.\ \bibinfo {pages}
  {333–342}\BibitemShut {NoStop}%
\bibitem [{\citenamefont {Gori}\ \emph {et~al.}(1998)\citenamefont {Gori},
  \citenamefont {Santarsiero}, \citenamefont {Vicalvi}, \citenamefont
  {Borghi},\ and\ \citenamefont {Guattari}}]{gori1998beam}%
  \BibitemOpen
  \bibfield  {author} {\bibinfo {author} {\bibfnamefont {F.}~\bibnamefont
  {Gori}}, \bibinfo {author} {\bibfnamefont {M.}~\bibnamefont {Santarsiero}},
  \bibinfo {author} {\bibfnamefont {S.}~\bibnamefont {Vicalvi}}, \bibinfo
  {author} {\bibfnamefont {R.}~\bibnamefont {Borghi}},\ and\ \bibinfo {author}
  {\bibfnamefont {G.}~\bibnamefont {Guattari}},\ }\bibfield  {title} {\bibinfo
  {title} {Beam coherence-polarization matrix},\ }\href
  {https://doi.org/10.1088/0963-9659/7/5/004} {\bibfield  {journal} {\bibinfo
  {journal} {Pure and Applied Optics: Journal of the European Optical Society
  Part A}\ }\textbf {\bibinfo {volume} {7}},\ \bibinfo {pages} {941} (\bibinfo
  {year} {1998})}\BibitemShut {NoStop}%
\bibitem [{\citenamefont {Pires}\ \emph {et~al.}(2021)\citenamefont {Pires},
  \citenamefont {Litchinitser},\ and\ \citenamefont {{a}o}}]{Pires:21}%
  \BibitemOpen
  \bibfield  {author} {\bibinfo {author} {\bibfnamefont {D.~G.}\ \bibnamefont
  {Pires}}, \bibinfo {author} {\bibfnamefont {N.~M.}\ \bibnamefont
  {Litchinitser}},\ and\ \bibinfo {author} {\bibfnamefont {P.~A.~B.}\
  \bibnamefont {{a}o}},\ }\bibfield  {title} {\bibinfo {title} {Scattering of
  partially coherent vortex beams by a pt-symmetric dipole},\ }\href
  {https://doi.org/10.1364/OE.427385} {\bibfield  {journal} {\bibinfo
  {journal} {Opt. Express}\ }\textbf {\bibinfo {volume} {29}},\ \bibinfo
  {pages} {15576} (\bibinfo {year} {2021})}\BibitemShut {NoStop}%
\bibitem [{\citenamefont {Agarwal}(2012)}]{agarwal2012quantum}%
  \BibitemOpen
  \bibfield  {author} {\bibinfo {author} {\bibfnamefont {G.~S.}\ \bibnamefont
  {Agarwal}},\ }\href@noop {} {\emph {\bibinfo {title} {Quantum optics}}}\
  (\bibinfo  {publisher} {Cambridge University Press},\ \bibinfo {year}
  {2012})\BibitemShut {NoStop}%
\bibitem [{\citenamefont {F{\"o}lling}\ \emph {et~al.}(2005)\citenamefont
  {F{\"o}lling}, \citenamefont {Gerbier}, \citenamefont {Widera}, \citenamefont
  {Mandel}, \citenamefont {Gericke},\ and\ \citenamefont
  {Bloch}}]{FollingNature}%
  \BibitemOpen
  \bibfield  {author} {\bibinfo {author} {\bibfnamefont {S.}~\bibnamefont
  {F{\"o}lling}}, \bibinfo {author} {\bibfnamefont {F.}~\bibnamefont
  {Gerbier}}, \bibinfo {author} {\bibfnamefont {A.}~\bibnamefont {Widera}},
  \bibinfo {author} {\bibfnamefont {O.}~\bibnamefont {Mandel}}, \bibinfo
  {author} {\bibfnamefont {T.}~\bibnamefont {Gericke}},\ and\ \bibinfo {author}
  {\bibfnamefont {I.}~\bibnamefont {Bloch}},\ }\bibfield  {title} {\bibinfo
  {title} {Spatial quantum noise interferometry in expanding ultracold atom
  clouds},\ }\href {https://doi.org/10.1038/nature03500} {\bibfield  {journal}
  {\bibinfo  {journal} {Nature}\ }\textbf {\bibinfo {volume} {434}},\ \bibinfo
  {pages} {481} (\bibinfo {year} {2005})}\BibitemShut {NoStop}%
\bibitem [{\citenamefont {Gori}\ \emph {et~al.}(1999)\citenamefont {Gori},
  \citenamefont {Santarsiero}, \citenamefont {Borghi},\ and\ \citenamefont
  {Guattari}}]{gori1999}%
  \BibitemOpen
  \bibfield  {author} {\bibinfo {author} {\bibfnamefont {F.}~\bibnamefont
  {Gori}}, \bibinfo {author} {\bibfnamefont {M.}~\bibnamefont {Santarsiero}},
  \bibinfo {author} {\bibfnamefont {R.}~\bibnamefont {Borghi}},\ and\ \bibinfo
  {author} {\bibfnamefont {G.}~\bibnamefont {Guattari}},\ }\bibfield  {title}
  {\bibinfo {title} {The irradiance of partially polarized beams in a scalar
  treatment},\ }\href
  {https://doi.org/https://doi.org/10.1016/S0030-4018(99)00130-3} {\bibfield
  {journal} {\bibinfo  {journal} {Opt. Commun.}\ }\textbf {\bibinfo {volume}
  {163}},\ \bibinfo {pages} {159} (\bibinfo {year} {1999})}\BibitemShut
  {NoStop}%
\bibitem [{\citenamefont {Goodman}(2008)}]{goodman2008introduction}%
  \BibitemOpen
  \bibfield  {author} {\bibinfo {author} {\bibfnamefont {J.~W.}\ \bibnamefont
  {Goodman}},\ }\href@noop {} {\emph {\bibinfo {title} {Introduction to Fourier
  optics. 2005}}}\ (\bibinfo {year} {2008})\BibitemShut {NoStop}%
\end{thebibliography}%

\newpage 

\onecolumngrid

\renewcommand{\thefigure}{S\arabic{figure}}
\setcounter{figure}{0}
\renewcommand{\thetable}{S\arabic{table}}
\setcounter{table}{0}
\renewcommand{\theequation}{S.\arabic{equation}}
\setcounter{equation}{0}

\section*{Supplementary Material}
In this supplementary material we present: (i) the explicit derivation of the van Cittert-Zernike theorem for polarized two-photon fields; (ii) explicit derivation of Eq. (9) of the main manuscript, and (iii) each element of the BCP matrix upon propagation to the far field.

\noindent \textbf{\large{1. Derivation of the van Cittert-Zernike theorem for polarized two-photon fields}}
\vspace{3mm}

\noindent Let us start by considering the second-order coherence matrix for a polarized, quasi-monochromatic field \cite{gori2000use}
\begin{equation}\label{Eq:matrix1}
J\left(\boldsymbol{r}_{1}, \boldsymbol{r}_{2}, z\right)=\left\langle j\left(\boldsymbol{r}_{1}, \boldsymbol{r}_{2}, z\right)\right\rangle=\left[\begin{array}{ll}
J_{H H}\left(\boldsymbol{r}_{1}, \boldsymbol{r}_{2}, z\right) & J_{H V}\left(\boldsymbol{r}_{1}, \boldsymbol{r}_{2}, z\right) \\
J_{V H}\left(\boldsymbol{r}_{1}, \boldsymbol{r}_{2}, z\right) & J_{V V}\left(\boldsymbol{r}_{1}, \boldsymbol{r}_{2}, z\right)
\end{array}\right],
\end{equation}
where $z$ stands for the propagation distance of the field and $\boldsymbol{r}$ is used to specify the position of a point at the transverse plane of observation. The elements of the coherence-polarization matrix [Eq. (\ref{Eq:matrix1})] are given by
\begin{equation}
J_{\alpha \beta}\left(\boldsymbol{r}_{1}, \boldsymbol{r}_{2}, z\right)=\left\langle j_{\alpha \beta}\left(\boldsymbol{r}_{1}, \boldsymbol{r}_{2}, z\right)\right\rangle=\left\langle E_{\alpha}^{(-)}\left(\boldsymbol{r}_{1}, z ; t\right) E_{\beta}^{(+)}\left(\boldsymbol{r}_{2}, z ; t\right)\right\rangle,
\end{equation}
with $\alpha,\beta = H,V$. The angle brackets denote time average, whereas the quantities $E^{(-)}_{\alpha}\pare{\boldsymbol{r},z;t}$ and $E^{(+)}_{\alpha}\pare{\boldsymbol{r},z;t}$ represent the negative- and positive-frequency components of the $\alpha$-polarized field-operator at the space-time point $\pare{\boldsymbol{r},z;t}$, respectively.

Note that Eq. (\ref{Eq:matrix1}) is valid for classical fields, as well as single-photon sources \cite{gori2000use,Saleh05PRL}. However, to include multi-photon effects higher-order correlation functions are needed. In particular, for light in an arbitrary quantum state, the polarized, two-photon four-point correlation matrix can readily be written as
\begin{equation}\label{Eq:matrix2}
G\left(\boldsymbol{r}_{1}, \boldsymbol{r}_{2} ; \boldsymbol{r}_{3}, \boldsymbol{r}_{4}, z\right)=\left\langle j\left(\boldsymbol{r}_{1}, \boldsymbol{r}_{2}, z\right) \otimes j\left(\boldsymbol{r}_{3}, \boldsymbol{r}_{4}, z\right)\right\rangle,
\end{equation}
\\
where $\otimes$ stands for the Kronecker (tensor) product. Note that the elements defined by the matrix in Eq. (\ref{Eq:matrix2}) are given by the four-point correlation matrix \cite{perez2017two}
\begin{equation}\label{Eqn:Prop}
G_{\alpha \beta \alpha^{\prime} \beta^{\prime}}\left(\boldsymbol{r}_{1}, \boldsymbol{r}_{2} ; \boldsymbol{r}_{3}, \boldsymbol{r}_{4}, z\right)=\left\langle E_{\alpha}^{(-)}\left(\boldsymbol{r}_{1}, z ; t\right) E_{\beta}^{(+)}\left(\boldsymbol{r}_{2}, z ; t\right) E_{\alpha^{\prime}}^{(-)}\left(\boldsymbol{r}_{3}, z ; t\right) E_{\beta^{\prime}}^{(+)}\left(\boldsymbol{r}_{4}, z ; t\right)\right\rangle,
\end{equation}
with $\alpha,\beta, \alpha', \beta' = H,V$. Furthermore, we realize that the elements defined by the previous equation follow a propagation formula of the form \cite{gori1999}
\begin{equation}\label{Eq:twophoton}
\begin{gathered}
G_{\alpha \beta \alpha^{\prime} \beta^{\prime}}\left(\boldsymbol{r}_{1}, \boldsymbol{r}_{2} ; \boldsymbol{r}_{3}, \boldsymbol{r}_{4}, z\right)=\int\int\int\int G_{\alpha \beta \alpha^{\prime} \beta^{\prime}}\left(\boldsymbol{\rho}_{1}, \boldsymbol{\rho}_{2} ; \boldsymbol{\rho}_{3}, \boldsymbol{\rho}_{4}, 0\right) K^{*}\left(\boldsymbol{r}_{1}, \boldsymbol{\rho}_{1}, z\right) K\left(\boldsymbol{r}_{2}, \boldsymbol{\rho}_{2}, z\right) \\
\times K^{*}\left(\boldsymbol{r}_{3}, \boldsymbol{\rho}_{3}, z\right) K\left(\boldsymbol{r}_{4}, \boldsymbol{\rho}_{4}, z\right) d^{2} \boldsymbol{\rho}_{1} d^{2} \boldsymbol{\rho}_{2} d^{2} \boldsymbol{\rho}_{3} d^{2} \boldsymbol{\rho}_{4},
\end{gathered}
\end{equation}
with the Fresnel propagation kernel defined by \cite{goodman2008introduction}
\begin{equation}
K(\boldsymbol{r}, \boldsymbol{\rho}, z)=\frac{-i \exp (i k z)}{\lambda z} \exp \left[\frac{i k}{2 z}(\boldsymbol{r}-\boldsymbol{\rho})^{2}\right],
\end{equation}
where $k=2\pi / \lambda$. Interestingly, in the context of the scalar theory, a spatially incoherent source is characterized by means of a delta-correlated intensity function, which indicates that subfields---making up for the whole source---at any two distinct points across the source plane are uncorrelated \cite{mandel1995optical}. In the same spirit, and following previous authors \cite{gori2000use,Saleh05PRL}, we define a partially polarized, spatially incoherent source as one whose four-point correlation matrix elements have the form
\begin{equation}\label{Eq:source}
G_{\alpha \beta \alpha^{\prime} \beta^{\prime}}\left(\boldsymbol{\rho}_{1}, \boldsymbol{\rho}_{2} ; \boldsymbol{\rho}_{3}, \boldsymbol{\rho}_{4}, 0\right)=\lambda^{4} I_{\alpha \beta}\left(\boldsymbol{\rho}_{1}\right) I_{\alpha^{\prime} \beta^{\prime}}\left(\boldsymbol{\rho}_{3}\right)\left[\delta\left(\boldsymbol{\rho}_{2}-\boldsymbol{\rho}_{1}\right) \delta\left(\boldsymbol{\rho}_{3}-\boldsymbol{\rho}_{4}\right)+\delta\left(\boldsymbol{\rho}_{2}-\boldsymbol{\rho}_{3}\right) \delta\left(\boldsymbol{\rho}_{1}-\boldsymbol{\rho}_{4}\right)\right].
\end{equation}
Here, $I_{\alpha\beta}\pare{\boldsymbol{\rho}}$ stands for the intensity, position-dependent, polarized two-photon source function. Note that the sum of delta functions in Eq. (\ref{Eq:source}) is a result of the wavefunction symmetrization due to photon (in)distinguishability \cite{perez2017two}.
By substituting Eq. (\ref{Eq:source}) into Eq. (\ref{Eq:twophoton}) we can thus obtain
\begin{equation}\label{Eq:VCZ}
\begin{aligned}
G_{\alpha \beta \alpha^{\prime} \beta^{\prime}}\left(\boldsymbol{r}_{1}, \boldsymbol{r}_{2} ; \boldsymbol{r}_{3}, \boldsymbol{r}_{4}, z\right)=& \frac{\exp \left[\frac{i k}{2 z}\left(r_{2}^{2}-r_{1}^{2}\right)\right] \exp \left[\frac{i k}{2 z}\left(r_{2}^{4}-r_{1}^{3}\right)\right]}{z^{4}} \iint I_{\alpha \beta}\left(\boldsymbol{\rho}_{1}\right) I_{\alpha^{\prime} \beta^{\prime}}\left(\boldsymbol{\rho}_{3}\right) \\
& \times \exp \left[\frac{-2 \pi i}{\lambda z} \boldsymbol{\rho}_{1} \cdot\left(\boldsymbol{r}_{2}-\boldsymbol{r}_{1}\right)\right] \exp \left[\frac{-2 \pi i}{\lambda z} \boldsymbol{\rho}_{3} \cdot\left(\boldsymbol{r}_{4}-\boldsymbol{r}_{3}\right)\right] d^{2} \boldsymbol{\rho}_{1} d^{2} \boldsymbol{\rho}_{3},
\end{aligned}
\end{equation}
which represents the extension of the van Cittert-Zernike theorem to two-photon, partially polarized fields.
\vspace{3mm}

\noindent \textbf{\large{2. Second-order coherence matrix}}
\vspace{3mm}

\noindent To show some consequences of the two-photon vectorial van Cittert-Zernike theorem, we now present an example where two-photon correlations are built up during propagation. Let us start from a spatially incoherent and unpolarized source, whose four-point correlation matrix is written, in the $\{\ket{HH},\ket{HV},\ket{VH},\ket{VV}\}$ basis, as 
\begin{equation} \label{Eq:Gini}
\begin{split}
G_{ini}\pare{x_1,x_2;x_3,x_4,0} & =  j_{ini}\pare{x_1,x_2,0} \otimes j_{ini}\pare{x_3,x_4,0} \\
& = \lambda^{4} I_{0}^{2}\cor{\delta\pare{x_{2}-x_{1}}\delta\pare{x_{3}-x_{4}} + \delta\pare{x_{2}-x_{3}}\delta\pare{x_{1}-x_{4}} } \\
& \hspace{5 mm} \times \begin{bmatrix}
1 & 0 & 0 & 0 \\
0 & 1 & 0 & 0 \\
0 & 0 & 1 & 0 \\
0 & 0 & 0 & 1
\end{bmatrix},
\end{split}
\end{equation}
where
\begin{equation}\label{Eq:jini}
j_{ini}\pare{x_1,x_2,0} = \lambda^{2}I_{0}\delta\pare{x_2 - x_1} \begin{bmatrix}
1 & 0 \\
0 & 1 
\end{bmatrix}
\end{equation}
stands for the two-point correlation matrix for a spatially incoherent and unpolarized photon source \cite{gori2000use}, and $I_{0}$ describes a constant-intensity factor. Note that, for the sake of simplicity, we have restricted ourselves to a one-dimensional case, i.e., we have taken only one element of the transversal vector $\boldsymbol{r} = (x,y)$.

To polarize the source, we make use of a linear polarizer. Specifically, we cover the source with a linear polarization grating whose angle between its transmission axis and the $x$-axis is a linear function of the form $\theta = \pi x/L$, with $L$ being the length of the grating. The four-point correlation matrix after the polarization grating can thus be written as
\begin{equation}\label{Eq:Gout}
\begin{split}
G_{out}\pare{x_1,x_2;x_3,x_4,0} & =  \cor{P^{\dagger}\pare{x_1}j_{ini}\pare{x_1,x_2,0}P\pare{x_2}} \otimes \cor{P^{\dagger}\pare{x_3}j_{ini}\pare{x_3,x_4,0}P\pare{x_4}} \\
& \times \text{rect}\pare{x_{1}/L}\text{rect}\pare{x_{2}/L}\text{rect}\pare{x_{3}/L}\text{rect}\pare{x_{4}/L} \\
& \times \lambda^{4} I_{0}^{2}\cor{\delta\pare{x_{2}-x_{1}}\delta\pare{x_{3}-x_{4}} + \delta\pare{x_{2}-x_{3}}\delta\pare{x_{1}-x_{4}} }
\end{split}
\end{equation} 
where the product of $\text{rect}\pare{\cdots}$ functions describe the finite size of the source, and the action of the polarization grating is given by the Jones matrix,
\begin{equation}\label{Eq:Jones}
P\pare{x} = \begin{bmatrix}
\cos^{2}\pare{\frac{\pi x}{L}} & \cos\pare{\pi x/p}\sin\pare{\pi x/L} \\
\cos\pare{\pi x/p}\sin\pare{\pi x/L}  & \sin^{2}\pare{\pi x/L} 
\end{bmatrix}.
\end{equation}

By substituting Eqs. (\ref{Eq:Gini})-(\ref{Eq:Jones}) into Eq. (\ref{Eq:VCZ}), we can obtain the explicit form of the polarized, four-point correlation matrix elements. As an example, we can find that, in the far-field---i.e. the region where the quadratic phase factor in front of the integral of Eq. (\ref{Eq:VCZ}) goes to one---the normalized four-point correlation function for H-polarized photons reads as
\begin{equation}
\begin{aligned}
G_{H H H H}\left(\nu_{1}, \nu_{2}, \nu_{3}, \nu_{4} ; z\right)=&\left[\operatorname{sinc}\left(\nu_{1}\right)+\frac{1}{2} \operatorname{sinc}\left(\nu_{1}-1\right)+\frac{1}{2} \operatorname{sinc}\left(\nu_{1}+1\right)\right] \\
& \times\left[\operatorname{sinc}\left(\nu_{2}\right)+\frac{1}{2} \operatorname{sinc}\left(\nu_{2}-1\right)+\frac{1}{2} \operatorname{sinc}\left(\nu_{2}+1\right)\right] \\
&+\frac{1}{16}\left[\operatorname{sinc}\left(2+\nu_{3}\right)\left(\operatorname{sinc}\left(\nu_{4}\right)+2 \operatorname{sinc}\left(1-\nu_{4}\right)+\operatorname{sinc}\left(2-\nu_{4}\right)\right)\right.\\
&+2 \operatorname{sinc}\left(1+\nu_{3}\right)\left(3 \operatorname{sinc}\left(\nu_{4}\right)+3 \operatorname{sinc}\left(1-\nu_{4}\right)+\operatorname{sinc}\left(2-\nu_{4}\right)+\operatorname{sinc}\left(1+\nu_{4}\right)\right) \\
&+\operatorname{sinc}\left(2-\nu_{3}\right)\left(\operatorname{sinc}\left(\nu_{4}\right)+2 \operatorname{sinc}\left(1+\nu_{4}\right)+\operatorname{sinc}\left(2+\nu_{4}\right)\right) \\
&+2 \operatorname{sinc}\left(1-\nu_{3}\right)\left(3 \operatorname{sinc}\left(\nu_{4}\right)+\operatorname{sinc}\left(1-\nu_{4}\right)+3 \operatorname{sinc}\left(1+\nu_{4}\right)+\operatorname{sinc}\left(2+\nu_{4}\right)\right) \\
&\left.+\operatorname{sinc}\left(\nu_{3}\right)\left(10 \operatorname{sinc}\left(\nu_{4}\right)+6 \operatorname{sinc}\left(1-\nu_{4}\right)+\operatorname{sinc}\left(2-\nu_{4}\right)+6 \operatorname{sinc}\left(1+\nu_{4}\right)+\operatorname{sinc}\left(2+\nu_{4}\right)\right)\right]
\end{aligned}
\end{equation}
with
\begin{equation}
\nu_1 = L\frac{x_2-x_3}{\lambda z}; \hspace{2mm} \nu_2 = L\frac{x_4-x_1}{\lambda z}; \hspace{2mm}  \nu_3 = L\frac{x_2-x_1}{\lambda z}; \hspace{2mm} \nu_4 = L\frac{x_4-x_3}{\lambda z}.
\end{equation}

Finally, by realizing that when monitoring the two-photon correlation function with two detectors, at the observation plane in $z$, we must set \cite{Saleh05PRL}: $x_2 = x_3$ and $x_1 = x_4$, we find that
\begin{equation}
G_{HHHH}\pare{\nu_1,\nu_2,\nu_3,\nu_4;z} = G_{HHHH}\pare{0,0,\nu_1,-\nu_1;z}.  
\end{equation} 
We can follow the same procedure as above to obtain the remaining terms of the four-point correlation matrix. 
\vspace{3mm}

\noindent \textbf{\large{3. BCP Matrix Elements}}
\vspace{3mm}

\noindent Each element of the final BCP matrix upon detection is given as follows:

\begin{equation}
\begin{aligned}
g^{(2)}_{\text{HHHH}}(\nu)&=\frac{1}{16} (10 \text{sinc}(\nu)^2+2 (6 \text{sinc}(\nu+1)+\text{sinc}(\nu+2)+6 \text{sinc}(1-\nu)+\text{sinc}(2-\nu)) \text{sinc}(\nu)\\
&+6 \text{sinc}(\nu+1)^2+\text{sinc}(\nu+2)^2+6 \text{sinc}(1-\nu)^2+\text{sinc}(2-\nu)^2\\
&+4 \text{sinc}(\nu+1) \text{sinc}(\nu+2)+4 (\text{sinc}(\nu+1)+\text{sinc}(2-\nu)) \text{sinc}(1-\nu)+16)
\end{aligned}
\end{equation}
\begin{equation}
\begin{aligned}
g^{(2)}_{\text{HHVV}}(\nu)&=g^{(2)}_{\text{VVHH}}(\nu)=\frac{1}{16} (2 \text{sinc}(\nu)^2-2 (\text{sinc}(\nu+2)+\text{sinc}(2-\nu)) \text{sinc}(\nu)+2 (\text{sinc}(1-\nu)\\
&-\text{sinc}(\nu+1))^2+\text{sinc}(\nu+2)^2+\text{sinc}(2-\nu)^2+16)
\end{aligned}
\end{equation}

\begin{equation}
\begin{aligned}
g^{(2)}_{\text{HHVH}}(\nu)&=g^{(2)}_{\text{VHHH}}(\nu)=-\frac{i}{16}  (\text{sinc}(\nu -2)^2+2 \text{sinc}(1 -  \nu ) \text{sinc}(  \nu -2)-\text{sinc}(  \nu +2)^2\\
&-2 \text{sinc}( \nu +2) \text{sinc}(  \nu +1 )+2 (\text{sinc}(1 -  \nu )-\text{sinc}(  \nu +1 )) (\text{sinc}(  \nu )+\text{sinc}(  \nu +1 )+\text{sinc}(1 - \nu )))\\
&=\left(g^{(2)}_{\text{HHHV}}(\nu)\right)^*=\left( g^{(2)}_{\text{HVHH}}(\nu)\right)^*
\end{aligned}
\end{equation}

\begin{equation}
\begin{aligned}
g^{(2)}_{\text{HVHV}}(\nu)&=g^{(2)}_{\text{VHVH}}(\nu)=\frac{1}{16} (2 \text{sinc}(\nu)^2-2 (\text{sinc}(\nu+2)+\text{sinc}(2-\nu)) \text{sinc}(\nu)\\
&+2 (\text{sinc}(1-\nu)-\text{sinc}(\nu+1))^2+\text{sinc}(\nu+2)^2+\text{sinc}(2-\nu)^2)
\end{aligned}
\end{equation}

\begin{equation}
\begin{aligned}
g^{(2)}_{\text{VHHV}}(\nu)&=g^{(2)}_{\text{HVVH}}(\nu)=\frac{1}{16} (6 \text{sinc}(\nu)^2-2 (\text{sinc}(\nu+2)+\text{sinc}(2-\nu)) \text{sinc}(\nu)+2 (\text{sinc}(1-\nu)\\
&-\text{sinc}(\nu+1))^2-\text{sinc}(\nu+2)^2-\text{sinc}(2-\nu)^2)
\end{aligned}
\end{equation}

\begin{equation}
\begin{aligned}
g^{(2)}_{\text{HVVV}}(\nu)&=g^{(2)}_{\text{VVHV}}(\nu)=\frac{i}{16}  (2 \text{sinc}(\nu+1)^2-2 \text{sinc}(\nu) \text{sinc}(\nu+1)-2 \text{sinc}(\nu+2) \text{sinc}(\nu+1)+\text{sinc}(\nu+2)^2\\
&-2 \text{sinc}(1-\nu)^2-\text{sinc}(2-\nu)^2+2 \text{sinc}(\nu) \text{sinc}(1-\nu)+2 \text{sinc}(1-\nu) \text{sinc}(2-\nu)\\
&=\left(g^{(2)}_{\text{VHVV}}(\nu)\right)^*=\left( g^{(2)}_{\text{VVVH}}(\nu)\right)^*
\end{aligned}
\end{equation}

\begin{equation}
\begin{aligned}
g^{(2)}_{\text{VVVV}}(\nu)&= \frac{1}{16} (10 \text{sinc}(\nu)^2+2 (-6 \text{sinc}(\nu+1)+\text{sinc}(\nu+2)-6 \text{sinc}(1-\nu)+\text{sinc}(2-\nu)) \text{sinc}(\nu)\\
&+6 \text{sinc}(\nu+1)^2+\text{sinc}(\nu+2)^2+6 \text{sinc}(1-\nu)^2+\text{sinc}(2-\nu)^2\\
&-4 \text{sinc}(\nu+1) \text{sinc}(\nu+2)+4 (\text{sinc}(\nu+1)-\text{sinc}(2-\nu)) \text{sinc}(1-\nu)+16)
\end{aligned}
\end{equation}

\vspace{3mm}

\textcolor{black}{\noindent \textbf{\large{4. Photon Statistics of Distinguishable (Polarized) Combined Fields}}}
\vspace{3mm}

\textcolor{black}{To describe the photon statistics evolution of the field that emerges from the polarization grating, we first note that once the unpolarized field traverses the polarization grating, two different field distributions are created, one that bears a horizontal polarization and one that is vertically polarized [see Fig. S1(a)]. We can then describe the propagation of these fields into the far-field by making use of Fresnel propagation [see Eq. (S.6)]. As one might expect, the spatial distribution of both independent and distinguishable fields (vertically- and horizontally-polarized) change upon propagation. This creates different intensity (mean photon number) patterns in the transverse planes located at different positions ($z$) along the propagation axis, see Figs. S1(b)-(i). Note that we have labeled as $z=0$ the field distribution immediately after the polarization grating.}    

\begin{figure*}[b!]
 \includegraphics[width=1.0\linewidth]{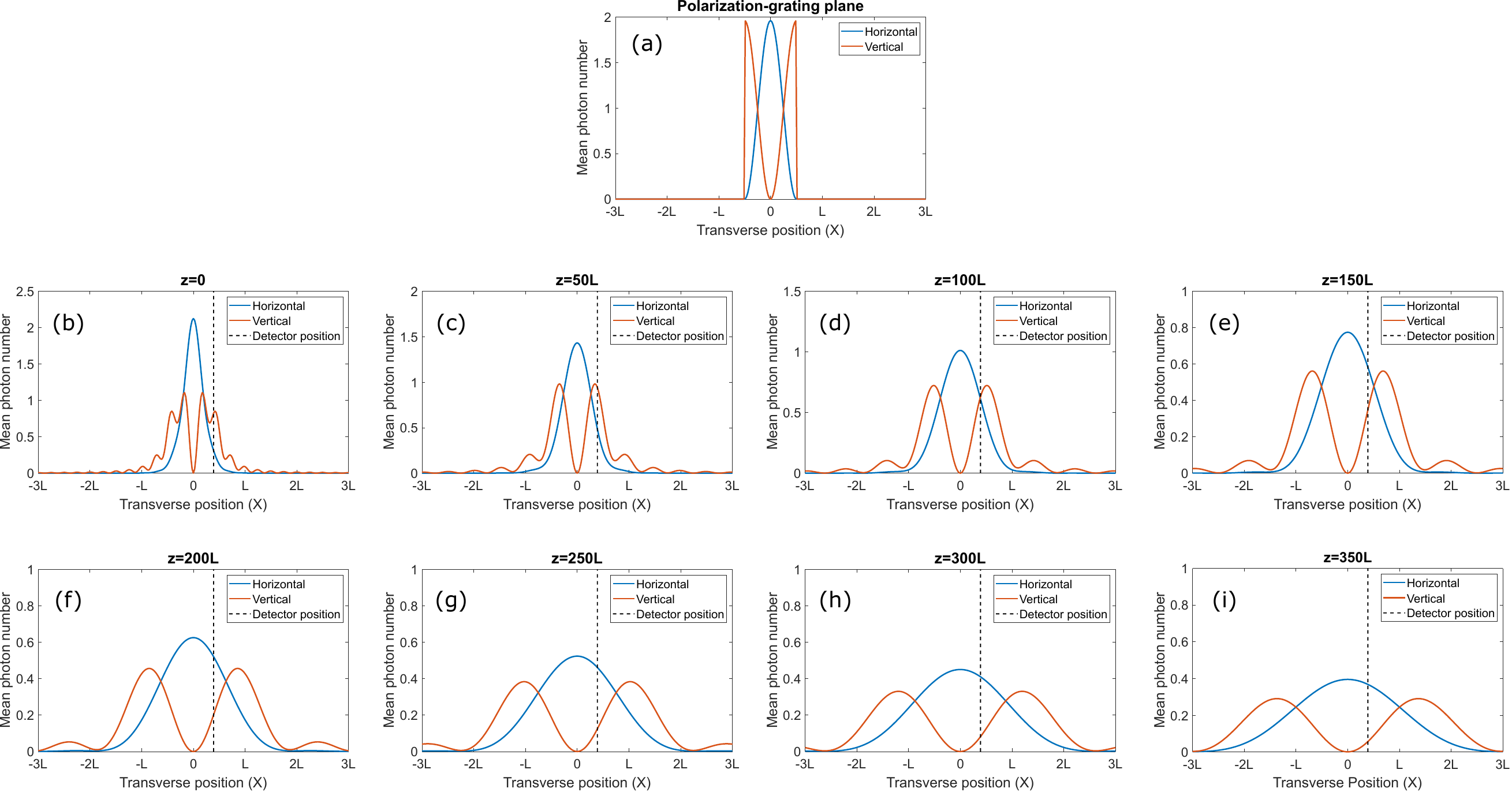}
 \caption{\textcolor{black}{(a) Spatial distribution of horizontally (blue line) and vertically (red line) polarized components of a thermal field at the polarization-grating plane. Each polarization component propagates in free-space following the Fresnel diffraction formula [see Eq. (6)], thus showing a different spatial field distribution at different positions along the propagation axis: (b) $z=0$, (c) $z=50L$, (d) $z=100L$, (e) $z=150L$, (f) $z=200L$, (g) $z=250L$, (h) $z=300L$, (i) $z=350L$. The photon distribution at each $z$ is obtained by placing a photon-number-resolving detector at $X=0.4 L$, depicted by the black-dashed line.}} 
 \label{fig:field_prop}
\end{figure*}

\textcolor{black}{To obtain the combined photon distribution, we first place a photon-number-resolving detector at an arbitrary position along the transverse plane. In the example shown in the main text, we selected a position $X=0.4L$ (depicted by the black-dashed line in Fig. S1), with $L$ being the length of the polarization grating. Following the theory of coherence of Glauber and Sudarshan. We can thus find that the photon distribution of the combined field, at a position $(X,z)$, is given by
\begin{equation}\label{joint_p}
    p^{(X,z)}(n) = \sum_{m=0}^{n}p^{(X,z)}_{H}(n-m)p^{(X,z)}_{V}(m),
\end{equation}
where $p^{(X,z)}_{H,V}$ stands for the photon distribution of the horizontally- and vertically-polarized fields, respectively. Because both fields are thermal, we can write their corresponding photon distribution as
\begin{equation}\label{thermal_p1}
    p^{(X,z)}_{H}(n) = \frac{\overline{n}_{H}(X,z)^{n}}{\pare{\overline{n}_{H}(X,z) + 1}^{n+1}}, 
\end{equation}
\begin{equation}\label{thermal_p2}
        p^{(X,z)}_{V}(n) = \frac{\overline{n}_{V}(X,z)^{n}}{\pare{\overline{n}_{V}(X,z) + 1}^{n+1}},
\end{equation}
with $\overline{n}_{H}(X,z)$ and $\overline{n}_{V}(X,z)$ being the mean photon number (extracted from the field distributions shown in Fig. S1) of the horizontal ($H$) and vertical ($V$) fields, at position $(X,z)$, respectively. Finally, by substituting Eqs. (\ref{thermal_p1}) and (\ref{thermal_p2}) into Eq. (\ref{joint_p}), with $X=0.4L$, we find the photon distributions shown in Fig. 3 of the main manuscript. Note that we can use Eq. (\ref{joint_p}) to evaluate the second-order quantum coherence, i.e. $g^{(2)}(\tau=0)$, at each point $(X,z)$, by realizing that $\ave{n(X,z)} = \sum_{n}^{\infty}np^{(X,z)}(n)$, and $\ave{n^{2}(X,z)} = \sum_{n}^{\infty}n^{2}p^{(X,z)}(n)$}.

\vspace{3mm}
\revA{\noindent \textbf{\large{5. Example Calculation: Indistinguishable Two-Photon States}}}
\vspace{3mm}

\begin{figure*}
 \includegraphics[width=0.8\linewidth]{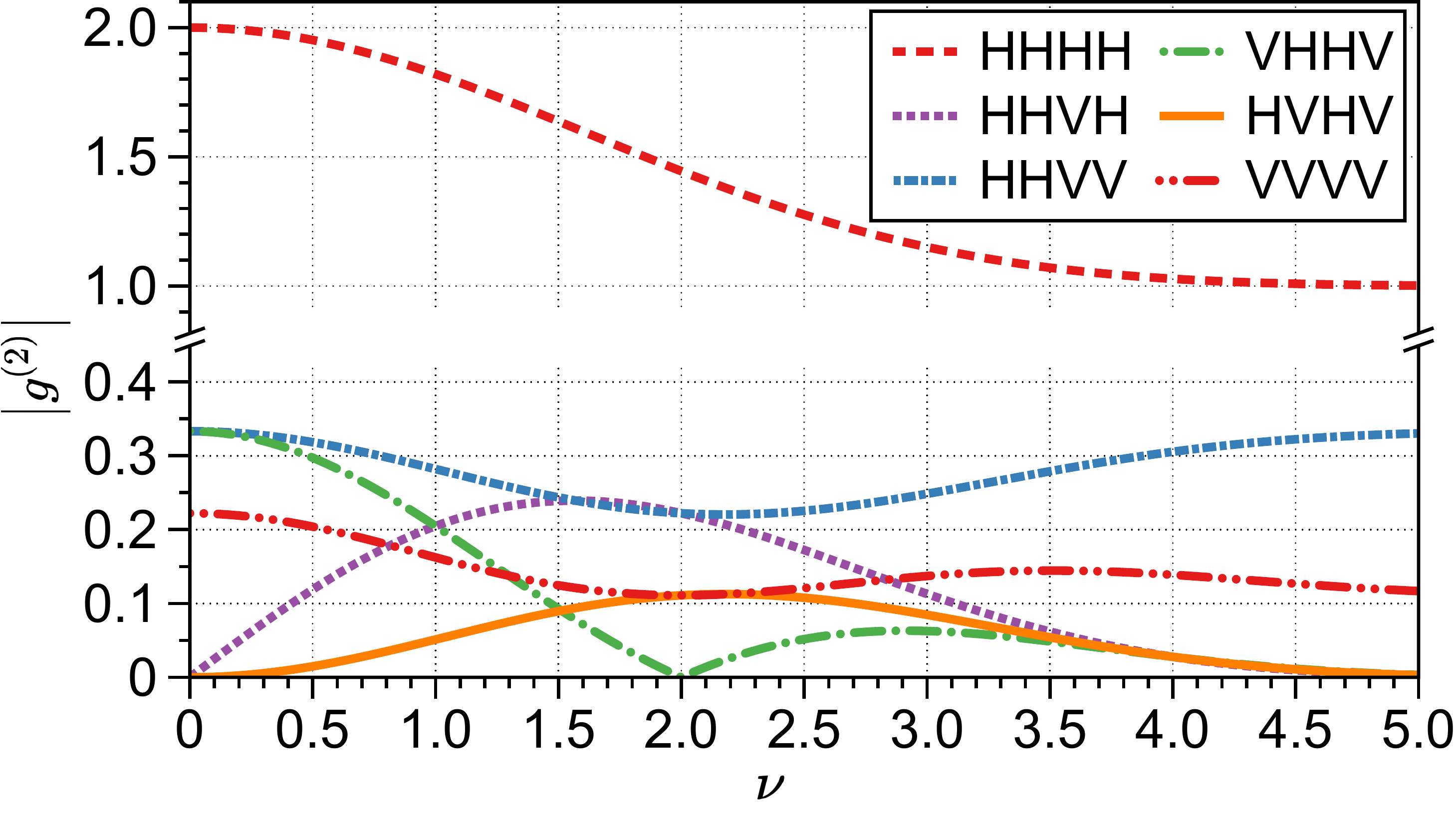}
 \caption{\revA{The behavior of various post-selected measurements for a system acted upon by a purely horizontal beam.  Here, all non-purely horizontal measurements exhibit sub-poissonian statistics. This is due to the vertical field is created purely by a Stern-Gerlach like mechanism through the repeated projective measurements by the polarization grating and the post-selection of our measurement.}} 
 \label{fig:ComponentsH}
\end{figure*}

\revA{In order to exemplify the characteristics of our system, we consider a two-photon indistinguishable state. Without loss of generality, we will restrict to indistinguishable horizontally-polarized photons. Here the beam coherence polarization matrix $j_{ini}\pare{x_1,x_2,0}$ can be written as}
\begin{equation}\label{Eq:jiniH}
j_{ini}\pare{x_1,x_2,0} = \lambda^{2}I_{0}\delta\pare{x_2 - x_1} \begin{bmatrix}
1 & 0 \\
0 & 0 
\end{bmatrix}.
\end{equation}
\revA{We can then propagate the system  using Eqs. (\ref{Eq:twophoton}) and (\ref{Eq:Gout})  and then normalize to find the finalized system. The final values of the system are in Figure \ref{fig:ComponentsH}. The vertical field and all correlations associated with it are created by the repeated projective measurements of the polarization grating and the post-selected measurements. This allows for a sub-poissonian fields to be created through  the use of only linear operations. Furthermore, two measurements will produce fields with spatial anti-bunching, the HVHV and HHVH measurement. \editor{It is important to note that this result holds for a multiphoton single-mode field as well.}}

\revA{The BCP matrix elements for the system are as follows:}

\begin{equation}
\begin{aligned}
g^{(2)}_{\text{HHHH}}(\nu)=\frac{1}{36} \left(6 \text{sinc}\left(\frac{  \nu }{2}\right)+4 \text{sinc}\left(\frac{  \nu }{2}+ \right)+\text{sinc}\left(\frac{  \nu }{2}+2  \right)+4 \text{sinc}\left( -\frac{  \nu }{2}\right)+\text{sinc}\left(2  -\frac{  \nu }{2}\right)\right)^2+1
\end{aligned}
\end{equation}
\begin{equation}
\begin{aligned}
g^{(2)}_{\text{HHVV}}(\nu)&=g^{(2)}_{\text{VVHH}}(\nu)=\frac{1}{3}-\frac{1}{36} \left(-2 \text{sinc}\left(\frac{  \nu }{2}+ \right)-\text{sinc}\left(\frac{  \nu }{2}+2  \right)+2 \text{sinc}\left( -\frac{  \nu }{2}\right)+\text{sinc}\left(2  -\frac{  \nu }{2}\right)\right)^2
\end{aligned}
\end{equation}

\begin{equation}
\begin{aligned}
g^{(2)}_{\text{HHVH}}(\nu)&=g^{(2)}_{\text{VHHH}}(\nu)=\frac{1}{36} i \left(-2 \text{sinc}\left(\frac{  \nu }{2}+ \right)-\text{sinc}\left(\frac{  \nu }{2}+2  \right)+2 \text{sinc}\left( -\frac{  \nu }{2}\right)+\text{sinc}\left(2  -\frac{  \nu }{2}\right)\right)\times\\
&\left(6 \text{sinc}\left(\frac{  \nu }{2}\right)+4 \text{sinc}\left(\frac{  \nu }{2}+ \right)+\text{sinc}\left(\frac{  \nu }{2}+2  \right)+4 \text{sinc}\left( -\frac{  \nu }{2}\right)+\text{sinc}\left(2  -\frac{  \nu }{2}\right)\right)\\
&=\left(g^{(2)}_{\text{HHHV}}(\nu)\right)^*=\left( g^{(2)}_{\text{HVHH}}(\nu)\right)^*
\end{aligned}
\end{equation}

\begin{equation}
\begin{aligned}
g^{(2)}_{\text{HVHV}}(\nu)&=g^{(2)}_{\text{VHVH}}(\nu)=-\frac{1}{36} \left(-2 \text{sinc}\left(\frac{  \nu }{2}+ \right)-\text{sinc}\left(\frac{  \nu }{2}+2  \right)+2 \text{sinc}\left( -\frac{  \nu }{2}\right)+\text{sinc}\left(2  -\frac{  \nu }{2}\right)\right)^2
\end{aligned}
\end{equation}

\begin{equation}
\begin{aligned}
g^{(2)}_{\text{VHHV}}(\nu)&=g^{(2)}_{\text{HVVH}}(\nu)=\frac{1}{36} \left(2 \text{sinc}\left(\frac{  \nu }{2}\right)-\text{sinc}\left(\frac{  \nu }{2}+2  \right)-\text{sinc}\left(2  -\frac{  \nu }{2}\right)\right)\times\\
&\left(6 \text{sinc}\left(\frac{  \nu }{2}\right)+4 \text{sinc}\left(\frac{  \nu }{2}+ \right)+\text{sinc}\left(\frac{  \nu }{2}+2  \right)+4 \text{sinc}\left( -\frac{  \nu }{2}\right)+\text{sinc}\left(2  -\frac{  \nu }{2}\right)\right)
\end{aligned}
\end{equation}

\begin{equation}
\begin{aligned}
g^{(2)}_{\text{HVVV}}(\nu)&=g^{(2)}_{\text{VVHV}}(\nu)=\frac{1}{36} i \left(2 \text{sinc}\left(\frac{  \nu }{2}\right)-\text{sinc}\left(\frac{  \nu }{2}+2  \right)-\text{sinc}\left(2  -\frac{  \nu }{2}\right)\right)\times\\
&\left(-2 \text{sinc}\left(\frac{  \nu }{2}+ \right)-\text{sinc}\left(\frac{  \nu }{2}+2  \right)+2 \text{sinc}\left( -\frac{  \nu }{2}\right)+\text{sinc}\left(2  -\frac{  \nu }{2}\right)\right)
\\
&=\left(g^{(2)}_{\text{VHVV}}(\nu)\right)^*=\left( g^{(2)}_{\text{VVVH}}(\nu)\right)^*
\end{aligned}
\end{equation}

\begin{equation}
\begin{aligned}
g^{(2)}_{\text{VVVV}}=\frac{1}{36} \left(\left(2 \text{sinc}\left(\frac{  \nu }{2}\right)-\text{sinc}\left(\frac{  \nu }{2}+2  \right)-\text{sinc}\left(2  -\frac{  \nu }{2}\right)\right)^2+4\right)
\end{aligned}
\end{equation}

\vspace{3mm}
\revA{\noindent \textbf{\large{\editor{5.1 Classical theory of coherence does not describe multiphoton scattering}}}}
\vspace{3mm}

\editor{In order to emphasize the quantum nature of our main result we present the classical analysis of the same system. We will present this calculation in the form of the Classical Beam Coherence Polarization (BCP) Matrix and highlight the differences that arise between the two analyses.  We begin by presenting the first order BCP-matrix for an incoherent, unpolarized source as
\begin{equation}\label{Eq:initalBCPClass}
J\left(\boldsymbol{r}_{1}, \boldsymbol{r}_{2}, z\right)=\left\langle j\left(\boldsymbol{r}_{1}, \boldsymbol{r}_{2}, z\right)\right\rangle=\left[\begin{array}{ll}
J_{H H}\left(\boldsymbol{r}_{1}, \boldsymbol{r}_{2}, z\right) & J_{H V}\left(\boldsymbol{r}_{1}, \boldsymbol{r}_{2}, z\right) \\
J_{V H}\left(\boldsymbol{r}_{1}, \boldsymbol{r}_{2}, z\right) & J_{V V}\left(\boldsymbol{r}_{1}, \boldsymbol{r}_{2}, z\right)
\end{array}\right]=\lambda^{2}I_{0}\delta\pare{x_2 - x_1} \begin{bmatrix}
1 & 0 \\
0 & 1 
\end{bmatrix}.
\end{equation}
Where in contrast to Eq. (\ref{Eq:matrix1}) we define $J_{\alpha\beta}=\langle E_\alpha^*(\boldsymbol{r}_1,z;t)E_\beta(\boldsymbol{r}_2,z;t)\rangle$. Therefore, the key difference between the classical approach to the system and the quantum approach previously outlined is that we propagate the first order classical correlations directly rather than converting it to the operator form and propagating the operators through space. This creates an immediate change when viewing the initial second order BCP matrix which is now written as
\begin{equation} \label{Eq:GiniClass}
\begin{split}
G_{ini}\pare{x_1,x_2;x_3,x_4,0} & =  j_{ini}\pare{x_1,x_2,0} \otimes j_{ini}\pare{x_3,x_4,0}\\
& = \lambda^{4} I_{0}^{2}\delta\pare{x_{2}-x_{1}}\delta\pare{x_{3}-x_{4}} 
\begin{bmatrix}
1 & 0 & 0 & 0 \\
0 & 1 & 0 & 0 \\
0 & 0 & 1 & 0 \\
0 & 0 & 0 & 1
\end{bmatrix},
\end{split}
\end{equation}
This removed the $\delta\pare{x_{2}-x_{3}}\delta\pare{x_{1}-x_{4}}$ term from Eq. (\ref{Eq:Gini})  which is ultimately the quantum term. In other words, a classical system presumes that there is no relation between individual photons regardless of measurements.  This effect holds when propagating through the polarization grating allowing the Van Cittert-Zernike theorem to be written as
\begin{equation}
\begin{aligned}
&G_{j k l m}^{(2)}(\mathbf{X},z)=\int d x_{1} \int d x_{2} \int d x_{3} \int d x_{4} C_{j k l m}(\boldsymbol{x}) F(\boldsymbol{x}, \mathbf{X},z) \delta\left(x_{1}-x_{2}\right) \delta\left(x_{3}-x_{4}\right),
\end{aligned}
\end{equation}
where the limits of integration for each integral is $-L/2$ to $L/2$, $\boldsymbol{x}=[x_1,x_2,x_3,x_4]$, $C_{j k l m}(\boldsymbol{x})$ is the coefficient of the $\mathbf{j}_{jk}\otimes\mathbf{j}_lm$  of the second order BCP matrix, and $j,k,l,m \in \{H,V\}$. Furthermore, $F\pare{\bm{x},\mathbf{X},z}$ is given as
\begin{equation}
\begin{aligned}
  F\pare{\bm{x},\mathbf{X},z}=\text{Exp}[\frac{2\pi i}{\lambda z}\pare{X_4x_4-X_3x_3+X_2x_2-X_1x_1}].
\end{aligned}
\end{equation}
Upon propagation we then have the far-field result as
\begin{equation}
    \mathbf{g}^{(2)}(\nu)=\begin{bmatrix}
1 & 0 & 0 & 0 \\
0 & 1 & 0 & 0 \\
0 & 0 & 1 & 0 \\
0 & 0 & 0 & 1
\end{bmatrix}
\end{equation}
Which vastly differs from the results presented in Fig. 2 of the main paper.  Mainly, we lose all spatial correlation as described before.  This results in our $g^{(2)}(0)$ diagonal elements to become 1 while all other elements are zero. This of course, matches the $g^{(2)}(0)$ obtained when modeling thermal light as a  classical source with a constant, non-fluctuating intensity, since bunching needs to be modeled by a quantum theory. Furthermore, the lack of off diagonal correlations erases the quantum degree of polarization we discovered in our paper and instead states that the source is unpolarized, as noted in the first order classical analysis of the system \cite{gori2000use}.  Overall, this highlights that our calculation produces results incapable of being replicated by a classical analysis.}

\end{document}